\documentstyle[12pt,aps,epsfig]{revtex} 
\begin{document}
\draft
\tighten
\title{Relativistic Heavy--Ion Collisions in \\
the Dynamical String--Parton Model}
\author{D. E. Malov, A. S. Umar, D. J. Ernst}
\address{Department of Physics and Astronomy, Vanderbilt University,
Nashville, TN 37235}
\author{D. J. Dean}
\address{Physics Division,
Oak Ridge National Laboratory, Oak Ridge, TN 37831--6373}
\date{\today}
\maketitle
\begin{abstract}
We develop and extend the
dynamical string parton model. This model, which is based on the salient features of QCD,
uses classical Nambu-Got\=o strings with the endpoints identified as partons, an 
invariant string breaking model of the hadronization process, and interactions described as
quark-quark interactions. In this work, the original 
model is extended to include a phenomenological quantization of the mass 
of the strings, an analytical technique for treating the incident 
nucleons as a distribution of string configurations determined by the experimentally measured 
structure function, the inclusion of the gluonic content of the nucleon through the 
introduction of purely gluonic strings, and the use of a hard parton-parton interaction taken 
from perturbative QCD combined with a phenomenological soft interaction. The limited number of 
parameters in the model are adjusted to $e^+e^-$ and $p\,$--$p$ data. Utilizing these 
parameters, the first calculations of the model for $p\,$--$A$ and $A$--$A$ collisions are 
presented and found to be in reasonable agreement with a broad set of data.
\end{abstract}
\pacs{25.75.+r, 12.38.Mh}

\section{Introduction}

Relativistic heavy-ion collisions at sufficiently high energies offer the
unique opportunity to probe highly excited dense nuclear matter under
controlled laboratory conditions. The ultimate objective \cite{QM99} of the
field of relativistic heavy-ion physics is to produce, detect, and study the
properties of the quark gluon plasma (QGP), a new state of matter
characterized by a transition to a phase of  deconfined quarks and gluons.
The fundamental confining property of the quantum chromodynamic (QCD) vacuum
means that the deconfined state of a QGP is not directly observable. What is
observable are hadronic and leptonic residues produced from the transient QGP
state. In order to connect the properties of a  QGP with experimental data
from real nuclear collisions, theoretical models of these very complex,
non-equilibrium collisions are needed. It is important that a variety of
models be developed as no single model can hope to be complete in its ability
to describe the system; each will have its strengths and limitations. We here
develop and present results from the dynamical string-parton (DSP) model, a
model which has as one of its strengths the detailed description of the
interaction in time, from the initial state of two colliding nuclei through
until the freeze-out of all the produced final-state particles.

The string-parton model\cite{DEAN's,Big,LNS,HM} is a fully dynamical but
classical string picture of hadrons and their collisions. The model
incorporates both confinement and asymptotic freedom. In this model, the
classical equations of motion for covariant strings determines the  motion of
hadronic matter according to the Lagrangian proposed by Nambu and Got\=o
\cite{artru74}. The string is viewed as a gluon field with massless quarks
attached at its endpoints. Mesons have a quark at one end and an antiquark at
the other, while baryons have a diquark at one end and a quark at the other.
Colliding strings interact via quark and gluon interactions. A covariant and
causal hadronization mechanism for the strings is included in the form of a
dynamic, probabilistic, and invariant  scheme for the decay of excited
strings \cite{Big}. The distinction of this model from others is the
inclusion of both real-time dynamical string evolution as well as the
inclusion of quark and gluon scattering as the sole source of the hadronic
interactions. The model is minimalist and completely dynamic. Once the
parameters of the string breaking and the quark and gluon interactions are
fixed, collisions are studied by evolving the string equations of motion in 
three--space and one--time dimensions. The model produces quite well
the average properties of a broad scope of reactions, including $e^+e^-$,
hadron-hadron \cite{Big}, lepton-nucleus \cite{LNS}, and nucleus-nucleus
collisions\cite{HM}.

The string-parton model is different in some essential ways from models based
on the ideas of the classical string that use collision generator concepts,
the  (LUND~\cite{lund}, FRITIOF~\cite{fritiof}, RQMD~\cite{rqmd}, and
QGSM~\cite{qgsm}) models. The strings in the DSP model are not longitudinal;
they are evolved according to the dynamics of the underlying Lagrangian. In
comparison to other approaches, the string parton model is also quite
minimalist. Fragmentation in most models is a process that involves a variety
of terms each with its own flavor-dependent parameters. The decay scheme
developed here is much simpler and flavor independent; flavor enters through
the quantization \cite{DM98} of the final-state hadronic masses. The
differences between the models are motivated by the intended applications of
the model. Since the DSP model describes reactions by calculating explicitly
the time evolution of the participant strings, we are limited in the
complexity of the fragmentation model that we are able to employ. In the DSP
model, the few parameters needed to describe fragmentation are determined from
$e^+e^-$ reactions \cite{DM98}. Quark-quark, quark-gluon, and gluon-gluon
interactions are introduced and adjusted to fit the basic features of
$p\,$--$p$ collisions.  At the $p\,$--$A$ and nucleus-nucleus level, the model
is completely predictive.

In the next section, we review the original model and some already published
refinements that we have made to that model. In the original model, all mesons
were treated as one generic meson of mass about 240 MeV. The model was not
capable of addressing data in which the final state mesons were specifically
identified. A technique has been developed to quantize the masses of the
mesons \cite{DM98}. This together with a review of how the string
configurations which make up the nucleon are determined from the
experimentally measured structure function \cite{DM99} is presented in Section
II. In section III the model is further developed by 1) extending the model of
the nucleon to include gluonic strings, thus incorporating the experimentally
measured gluonic content of the nucleon into the model, and 2) extending the
quark-quark, quark-gluon, and gluon-gluon interactions to contain both a hard
process taken from perturbative QCD and a soft process that is determined
phenomenologically. Also in Section III, we determine some of the limited
number of  parameters in the model from $e^+e^-$ data. In Section IV, we
examine $p\,$--$p$ collisions where we find further constraints on the
parameters and demonstrate the ability of the model to describe hadron-hadron
collisions. In Section V we present the first results for $p\,$--$A$ collisions
of the new model, and in Section VI give the first results for $A$--$A$
collisions. Finally we summarize what we have learned from the model and
provide thoughts on future work, both applications and possible
improvements and generalizations.

\section{New Enhanced Model}

In this section we review the model and several enhancements to the model ---
these include  a new analytic method \cite{DM99} for determining the 
distribution of string solutions that make up a nucleon directly from the
experimentally measured structure function and the method \cite{DM99}  for
quantizing the masses of the mesons. These have been published elsewhere. 
The first results for hadronic collisions which employ these enhancements are
presented here in Sections III, IV, and V.

The classical Nambu-Got\=o Lagrangian \cite{artru74} provides a covariant 
description of the motion of a classical
 string with massless endpoints. The Lagrangian is defined as 
proportional to the invariant  area swept out by the string,  
\begin{equation} 
S=-\kappa\,\int\,dA=-\kappa\,\int_{\tau_i}^{\tau_f}\,\int_0^{\sigma_f}\,
\left\{ \left( \frac{\partial x^\mu}{\partial \sigma}\,\frac{\partial x_\mu}
{\partial 
\tau} \right) ^2 
- \left( \frac{\partial x}{\delta\sigma}\right)^2\, \left(
\frac{\partial x}{\partial
\tau} \right) ^2 \right\}^{\frac{1}{2}}\;,
\label{lagr} 
\end{equation}
where $\sigma$ and $\tau$ are parametric variables that determine the location
of the point $x^\mu (\sigma ,\tau )$, where  $x^\mu (\sigma ,\tau )$ is a
point which lies on the area swept out by the string. The usual continuum
mechanics algebra produces the equations of motion for the string. The
proportionality constant $\kappa$ is a parameter of the model and has units of
energy/length. It is called the string constant and determines the scale at
which the space-time evolution occurs. It is a parameter of the model. Lattice
gauge calculations \cite{bali} produce the linear potential between quarks
implied by this Lagrangian. For strings with total momentum zero, the
solutions to the Lagrangian are characterized by a closed trajectory. For each
trajectory, the endpoints of the string travel around the closed trajectory at
the speed of light.  The location of the string connecting the endpoints
follows from the equations of motion. To obtain strings with non-zero momenta,
the stationary strings are simply boosted relative to their rest frame.

For each trajectory, the endpoints of the string carry a finite and well
determined fraction of the energy and momentum of the string. We identify
these endpoints as quarks \cite{Big}. As the endpoints move at the speed of
light, they represent massless quarks. For mesons, the endpoints are taken to
be a quark and anti-quark; for nucleons, the endpoints are identified with a
quark and a di-quark. The electromagnetic structure function provides an
experimental measure of the longitudinal momentum fraction $x$ carried by the
quarks, with $x$ defined by
\begin{equation} 
x=\frac{k_0+k_3}{P_0+P_3}\,\,, 
\label{longmo} 
\end{equation}
where $k_0$ and $k_3$ ($P_0$ and $P_3$) are the energy and $z$-component of
the quark's (whole string's) momentum. For any string 
trajectory, the
momentum fraction carried by the quarks can be calculated by averaging over
the orientation of the string and  over both one complete period of
motion, and finally taking the limit to the infinite momentum frame. Each
trajectory thus implies a particular result for the structure function.
Turning this around,we use the experimentally measured structure function of a
hadron to determine a distribution of trajectories that will
reproduce the experimentally measured results. This concept of utilizing a
distribution of trajectories of classical strings to model the structure
function is used here and has also been applied in Ref.~\cite{fri98}. In
Ref.~\cite{DM99} this concept was pushed one step further. There it was shown
that the structure function can be analytically related to the distribution
of straight line segments making up the trajectories, assuming that
trajectories are polygons made up of straight line segments. The results of
these calculations demonstrate that the nucleon can be modeled as 
strings with a large
number of straight line segments --- a significant fraction of the distribution
is for strings with greater than ten segments. The nucleon has no probability
\cite{DM99} of being a one-segment or one-dimensional 
 solution. In this one-segment solution,
the quarks oscillate back and forth along a straight line, a
solution which is called the {\it yo-yo}. The structure function for the
yo-yo solution can be calculated analytically \cite{DM99,DMthesis}. The yo-yo
is the only shape which contributes in the limit $x\rightarrow 1$. For this
solution, and this solution only, the probability of one quark carrying all
the momentum goes to zero linearly in $1-x$. For the three-quark nucleon it
is not surprising that this solution is absent. For the two-quark pion,
however, there is a finite probability for the pion to be a yo-yo, or nearly
so. Many models, however, are based on or are motivated by the yo-yo solution
to the classical string model.

The results of Ref.~\cite{DM99} provides a means for initializing the proton as a
distribution of strings with the correct parton momentum distribution. The
analytical results are supplemented by a prescription which randomly attaches
the calculated distribution of straight line segments to form closed
trajectories. This collection of strings then forms the nucleon. A moving
nucleon results from boosting the collection of strings from
the rest frame. This provides the
needed model of an incident nucleon, and, by distributing nucleons according
to their experimentally measured spatial and momentum distribution in a
nucleus, a model of an incident nucleus. Note that the model of the
nucleon/nucleus is completely determined by experimental data 
other than the hadronic reactions which are to be studied.
We must also supplement the dynamics by interactions and decays.

A model of the hadronization process is necessary if we are to model
high-energy nuclear processes. The hadronization process is believed to take
place via soft non-perturbative mechanisms.
The DSP model assumes that an excited and therefore stretching 
color flux-tube
breaks via the creation of a $q\bar q$ pair with the color field of the new
pair equal and opposite that found in the parent string. A covariant description
\cite{Big} results from assuming that the probability of a string breaking is
proportional to the invariant area, $\delta A$, swept out by the string,
\begin{equation}
\delta {\cal P}=\Lambda\, \delta A\,\,.
\label{dprob}
\end{equation}
The proportionality constant $\Lambda$ is a parameter to be determined from
experimental data. It determines the rate at which strings break and thus will be
intimately related to the multiplicity of a reaction as well as
other experimentally
measurable  quantities. This decay law is invariant because it is defined in
terms of an invariant area, and it is causal as the breaking occurs at a
single point on the string. The probability that a string which is present at
time $t_1$ will survive until time $t_2$ is then the exponential
\begin{equation} 
{\cal P}(t_1\rightarrow t_2)=\exp(-\Lambda A(\delta t))\,\,,
\label{eprob} 
\end{equation}
where $\delta t= t_2-t_1$. 

Hadronization is then treated stochastically. Aa a  string evolves 
in time, it decays  according to the probability given by
Eq.~(\ref{eprob}). If a string is to decay, it breaks at a point chosen
randomly along its length. We have examined the possibility of calculating the
tension locally along the string and then having the string break at each
point according to the tension at that point. We found that
this possible refinement  had no significant impact on experimentally measurable
quantities, so we do not utilize it. 

Much of the qualitative features of hadronization
are contained in this simple model:
\begin{itemize}
\item The intuitive picture of a string breaking more often the more it is
stretched is built in. The more highly excited the string, the more likely it
will break since the string sweeps through more invariant area.
\item The uniform distribution of pions and the rapidity plateau emerge
naturally.
\item The leading particle effect results from having those particles created
in the central region arising from quark-quark interactions that produce
highly excited strings which fragment first. String pieces arising from the
primary quarks are less excited and fragment last.
\end{itemize}

This model presents a problem that is completely analogous to a problem with
the classical Bohr model of the atom. In the Bohr model, the electron would
continue to radiate until it spiraled into the nucleus. Here, a string would
continue to decay until it became an infinite number of very small pieces
each with very little energy. In the original DSP model
\cite{DEAN's,Big,LNS,HM}, a generic meson mass was introduced. Meson strings
were allowed to decay if they had sufficient energy to make two of these
generic mesons. Otherwise, the string could not decay and was
identified as a generic meson. For baryonic strings, strings with masses
greater than the nucleon mass plus the generic meson mass were permitted to
decay; the rest would be identified with the nucleon. This scheme is very
limiting. It allowed the model to address questions of mass flow, momentum
flow, and multiplicity, but could not address
meson identification, as is common for contemporary experimental data.

The model was generalized \cite{DM98} by quantizing the mass of the classical
strings. This is done by creating mass windows, energy regions that
extend from the physical meson mass, $m_i$, to the physical  mass plus a
small increment, $\delta m_i$. If the mass of a string created in a
hadronization process lands within one of these windows, the point at which
the string decays is adjusted a small amount to produce a section of string
with a mass equal to the physical meson mass. Once a string is identified
with a meson, it is no longer allowed to decay. Thus the mass of a string in
the final state is quantized to be the mass of a physical meson (within the
small width that results from the numerical discretization of the string).
Strings with a mass larger than the heaviest meson (we take a cutoff of 1
GeV) are always allowed to decay, as are strings with masses that fall
between the mass windows. Phase space arguments were provided that predict
the width of each window $\delta m_i$. These results were found \cite{DM98}
to be in qualitative agreement with results that were obtained from fitting
to meson multiplicities measured in $e^+e^-$ experiments, as will be described
in more detail in the next section.

A striking experimental feature of both the $e^+e^-\rightarrow q\bar
q\rightarrow$ hadrons reaction and the $pp\rightarrow$ hadrons reaction is
that the mean transverse momentum is always roughly $\langle p_T
\rangle\simeq0.35$ GeV/c. This leads to the assumption that the production of
low transverse momentum  ($p_T\le 1.5$ GeV/c) in these experiments is due
primarily to the fragmentation of the gluon field that binds the quarks. This
can be modeled by giving the quarks created in a string breaking equal and
opposite non-vanishing transverse momenta. Although the transverse
cross-section comprises a small part of the total cross-section, it is of
significant importance in relativistic heavy-ion collisions since it can be
used as a measure of the {\it stopping} experienced by the nuclei
\cite{Bj83}. Various models of $q\bar{q}$ pair production \cite{SPT} predict
a Gaussian transverse momentum distribution during the string breaking, but
this falls far short of the experimentally measured distributions. Extensions
of such models to infinitely long, uniform color-electric  flux-tubes with
finite transverse size may be a remedy \cite{SS90}. In the absence of any
fundamental calculations, we chose to parameterize the transverse momentum
acquired by the created quarks with a simple exponential distribution
function 
\begin{equation} 
f(p_T)p_T dp_T\propto e^{-\alpha p_T}p_T dp_T\;.
\label{ptd} 
\end{equation}
 The average experimental transverse momentum is approximately related to 
$\alpha$ in Eq.~(\ref{ptd}) by $\langle p_T\rangle=2/\alpha$. This 
is not an
analytically exact  relationship as the  transverse direction to the string 
is not,
particularly after a number of quark collisions, the same as the transverse 
direction in the reaction plane
and because the high-momentum tail in Eq.~(\ref{ptd}) cannot occur as these
momenta would violate conservation of energy-momentum.

The final ingredient needed to be able to model hadronic collisions is a
model of the hadron-hadron interaction. In this work, we further refine
the model of the hadronic interactions. This new model is developed in
Section IV. We also modify the model of the nucleon so as to include the
experimentally determined gluonic content of the nucleon into the model. This
development is also given in Section IV. Preliminary results which
contained some of these new elements developed here can be found in
Refs.~\cite{DMthesis,atomki}.

\section{Hadronization and electron-positron collisions}

In this section we describe our model of the hadronization process and how 
the parameters which control the process are determined from $e^+e^-$ data.
The simulation of an $e^+e^-$ collision begins at the $q\bar q$  vertex where a
quark and anti-quark are created with oppositely directed momenta and move
apart, stretching the string between them. The parameters that relate to  the
quark and gluon interactions are to be determined from $p\,$--$p$  collisions
which will be presented in the next section. However, we will also show here
that the quark and gluon interactions play an important role in $e^+e^-$
dynamics. This physics had been neglected in the previous version of the DSP
model.  The determination of the parameters of the model no longer separates
into first using $e^+e^-$ data to determine the hadronization parameters and
then using $p\,$--$p$ interactions to determine the interaction parameters.
Instead, the best interaction parameters must be used while determining the
hadronization parameters, and vice versa.

The first parameter is the string constant $\kappa$ that
sets the energy scale of the model. The Regge
model of highly excited hadrons \cite{perkins} pictures these states
as rotational states built on individual excited states. The string model also
provides a picture of excited states arising from rotating strings and gives
the linear trajectories found in the Regge model. The slope of these
trajectories is the string tension and is found experimentally \cite{perkins}
to be $\kappa=$ 0.88 GeV/fm. We will utilize this value; thus
this parameter is determined completely external to our model.

\begin{figure}[thbp]
%\vspace*{+2mm}
\begin{minipage}[t]{7.5cm}
{\epsfxsize=7cm\epsfysize=4.7cm \epsfbox{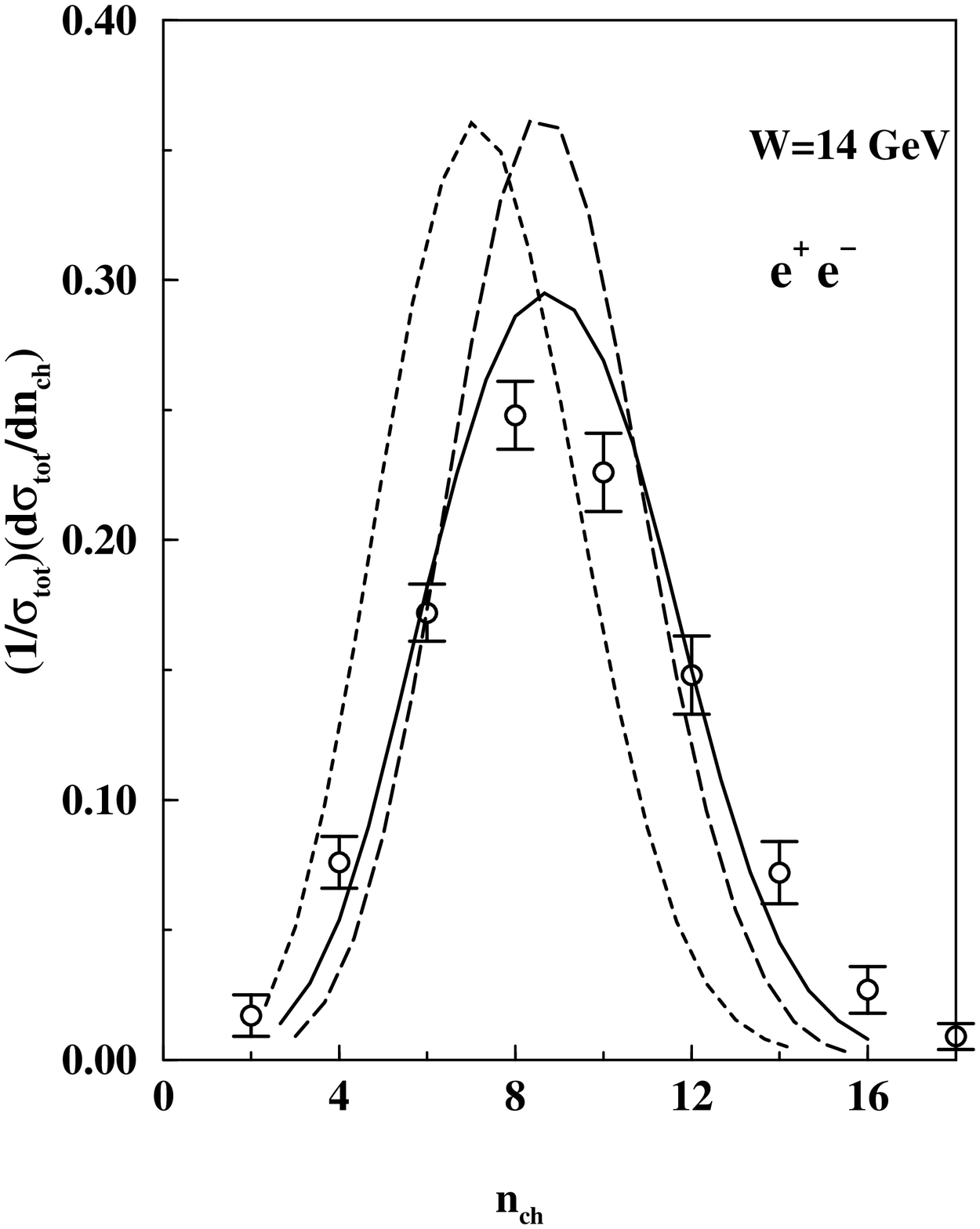}}
\caption[]{\footnotesize  Charged particle multiplicity distributions in  $e^+e^-$ collisions at
$\sqrt{s}=$ 14 GeV. The data are from the TASSO experiment
\protect{\cite{TASSO}}. The solid line is the prediction of the SP model with
the interaction of the quarks turned on and  $\Lambda =1$. The long-dashed
curve corresponds to $\Lambda =1$ and the interaction of the quarks turned off.
The short-dashed curve has the quark interaction turned on but with  
$\Lambda =4$.}
\label{fig1_multi}
\end{minipage} \hfill
\vspace*{-9.2cm}
 \hspace*{6.5cm} \hfill
\begin{minipage}[t]{7.5cm}
{\epsfxsize=7cm\epsfysize=4.7cm \epsfbox{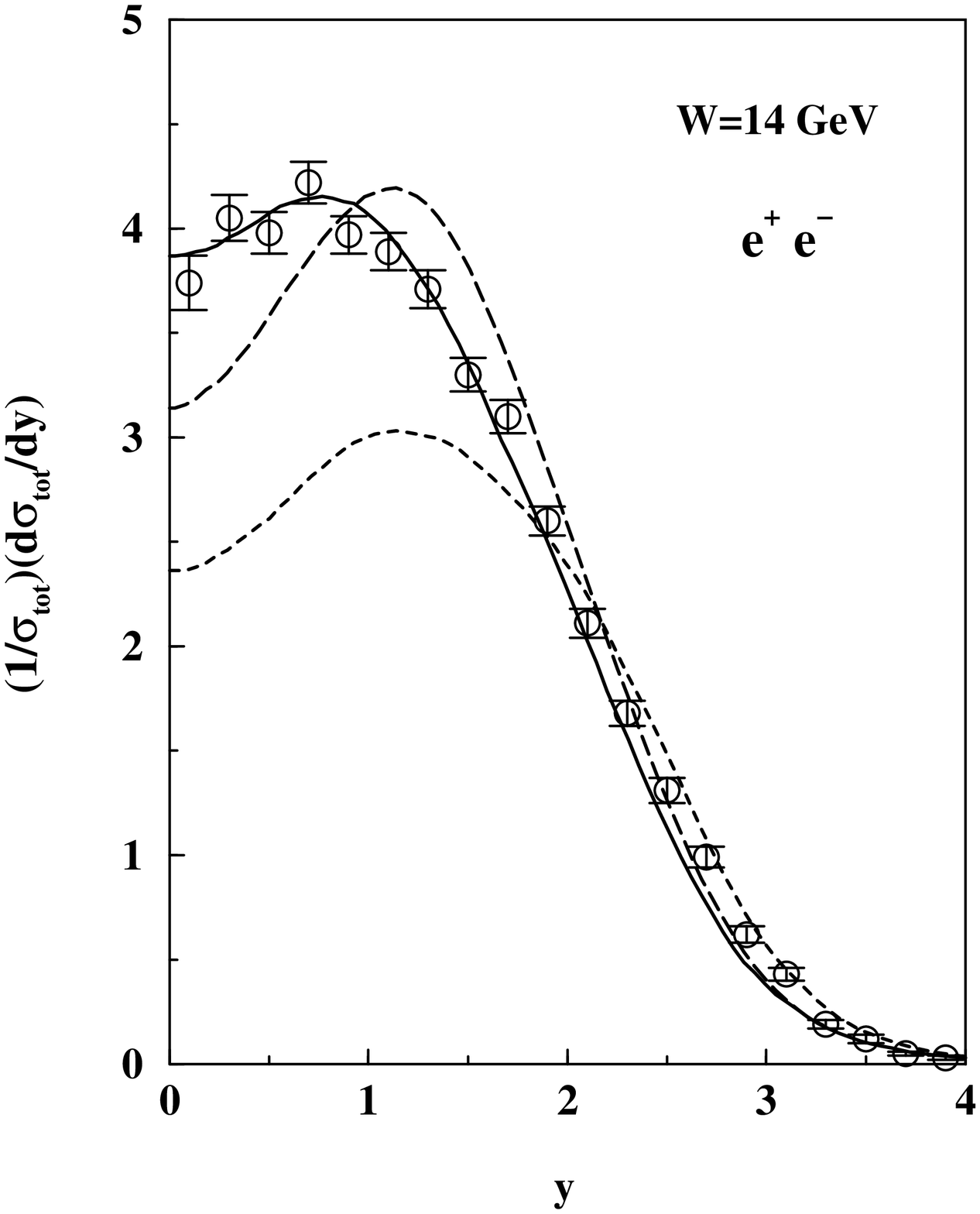}}
%\vspace*{2pt}
\caption[]{\footnotesize   The same as
Fig.~(\protect{\ref{fig1_multi}}) except the rapidity distribution is
presented.}
\label{fig1_rapid}
\end{minipage}
\end{figure}
\begin{figure}[thbp]
%\vspace*{+2mm}
\begin{minipage}[t]{7.5cm}
{\epsfxsize=7cm\epsfysize=4.7cm \epsfbox{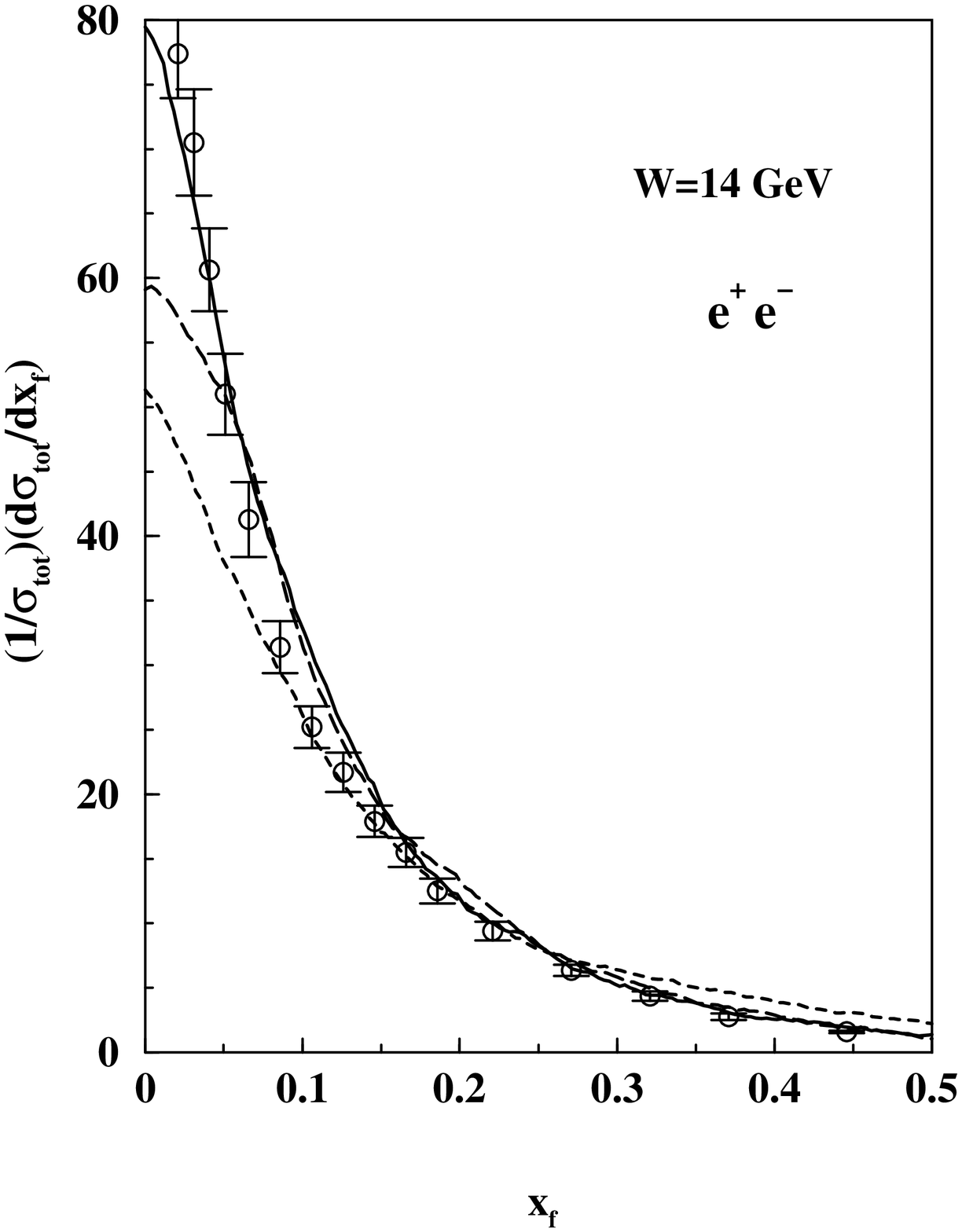}}
\caption[]{\footnotesize  The same
as Fig.~(\protect{\ref{fig1_multi}}) except the fractional longitudinal
momentum distribution is presented. }
\label{fig1_xf}
\end{minipage} \hfill
\vspace*{-6.1cm}
 \hspace*{6.5cm} \hfill
\begin{minipage}[t]{7.5cm}
{\epsfxsize=7cm\epsfysize=4.7cm \epsfbox{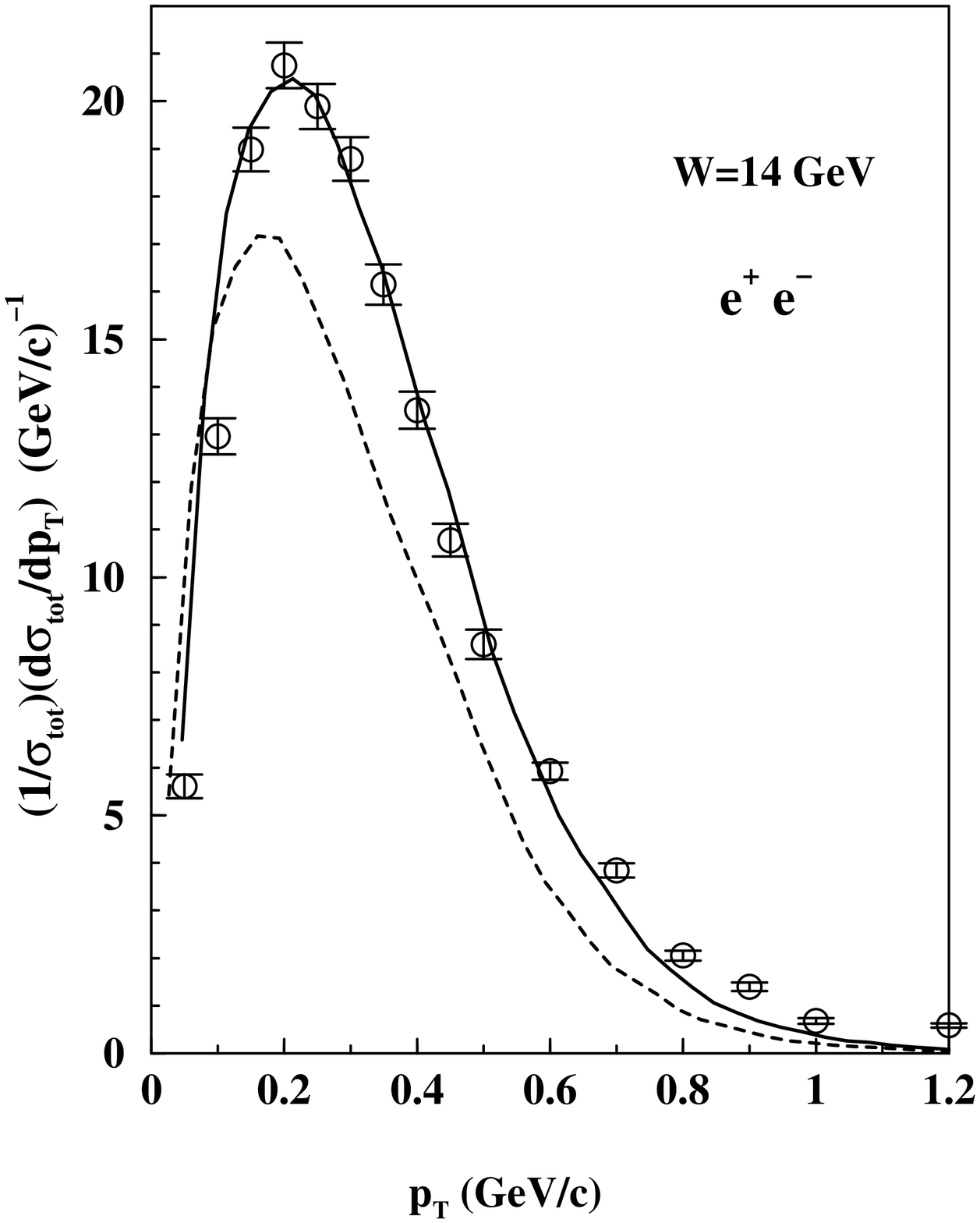}}
%\vspace*{2pt}
\caption[]{\footnotesize The same as Fig.~(\protect{\ref{fig1_multi}}) except the transverse
momentum distribution is presented. The curve with the quark interaction
turned off is not presented because it produces results very near to the solid
line.   }
\label{fig1_pt}
\end{minipage}
\end{figure}

 
\vspace*{3.8cm}
The next parameter to be determined is the decay parameter $\Lambda$ in
Eq.~(\ref{eprob}). This parameter determines the rate at which excited 
strings
decay and has a broad effect on measurable quantities in $e^+e^-$ reactions.
However, we have also discovered a new phenomena, the dynamics of the $e^+e^-$
reaction are also affected by the quark and gluon interactions. In the next
section we will describe, in detail, our model of these interactions. Here, we
will demonstrate their effect on the $e^+e^-$ theoretical results. To do
this,  we present calculations which utilize the best-fit parameters for the
interactions from the next section to compare with results in which we neglect
these interactions completely. In Fig.~(\ref{fig1_multi}) we show the charged
particle multiplicity  distribution for $e^+e^-$ collisions at 14 GeV. The
data are from the TASSO collaboration \cite{TASSO}. We compare the solid line,
the calculation with $\Lambda =$ 1 $fm^{-2}$ and the rescattering between the
created quarks included with the dashed line in which the rescattering is
turned off. The average multiplicity is not changed by the rescattering;  the
peaks of the two curves are at the same value of the multiplicity. However,
the rescattering broadens the curve and moves it much closer to the data. The
effect of increasing $\Lambda$ is to change the average as well as the width.
This is illustrated by the short-dashed curve in Fig.~(\ref{fig1_multi}).
In Fig.~(\ref{fig1_rapid}) the rapidity 
$y=(1/2)\log((E+p_l)/(E-p_l))$
distribution is shown, and in
Fig.~(\ref{fig1_xf}) the scaled longitudinal momentum $x_f=2p_l/\sqrt{s}$
distribution $(1/\sigma_{tot})\,(d\sigma/dp_f)$ is shown. The curves are coded
as in Fig.~(\ref{fig1_multi}). We have found that including the rescattering
of the created quarks is necessary if we are to fit these results. Notice that
the rapidity distribution without the rescattering, the dashed curve in
Fig.~(\ref{fig1_rapid}), has a peak between  $y= 1$ and $y= 2$. We have found
this peak to be a persistent feature in all calculations which do not include
this additional rescattering. Similarly, the rescattering naturally enhances
the fractional longitudinal momentum distribution near $x_f= 0$, as is needed
in the data.

By looking at the parameters and assumptions of our model, we can understand
why this additional scattering must be included.
Upon breaking, the newly created
endpoints of the string do {\it not} satisfy the energy-momentum relationship
of an on-shell massless particle. They are virtual and have a mass deficit. 
These quarks proceed along the string absorbing the energy in the string until
they have gained sufficient energy to come on the mass shell. In our model, the
created quarks are assigned a longitudinal momentum equal to the created
transverse momentum. This assignment is chosen as this will allow the quark to
come on to the mass shell as quickly as possible. We have examined other
assignments of longitudinal momentum during the string breaking, but the data
seems to prefer this original choice. A part of our model is not to allow the
quarks to interact until they are on-shell. This is  motivated by the view
that these interactions are mediated by gluonic exchanges. The propagator for
the exchanged gluon would be suppressed if one of the interacting particles is
highly virtual.  We model this by not allowing the off-shell quarks and gluons
to interact.

\vspace*{1.cm}

The average transverse momentum created in a string breaking is related to 
$\alpha$ in Eq.~(\ref{ptd}) approximately by $\langle p_T\rangle=2/\alpha$. We
will find that $\alpha=3.88\;(GeV/c)^{-1}$, a value determined by the
experimentally measured low transverse momentum distributions in $e^+e^-$
reactions. The time to come on shell is proportional to $p_T$ in the rest
frame of the created pair. The quarks will be  of the order of $2\langle p_T
\rangle$ apart when they come on shell, or about 0.70 GeV/c. This is roughly
equal to the interaction radius $r_g$ for the quarks or gluons to interact.
Thus one would expect a significant fraction of the created quark pairs to
interact with each other after their creation.

Another way of viewing this is that the assignment of a transverse momentum
(and an accompanying longitudinal momentum) at the string breaking is,
itself, a model of the interaction of the created quark pairs. However, this
model does not produce the experimental data. Without adding additional
parameters to the model, but simply by making it more internally consistent, we
find improved results, particularly for rapidity and
$x_f$ distributions. We note that
the quark-quark, quark-gluon, and gluon-gluon interactions will be found to
have  hard (high-momentum)  and  soft (low-momentum) contributions. The value
of $\alpha$ restricts the momenta of the created pair to a region where only
the soft amplitude is effective in this rescattering.

In Figs.~(\ref{fig1_multi}-\ref{fig1_xf}), we also show results with $\Lambda
= 4$ fm$^{-2}$ (short-dashed curves) to compare with $\Lambda= 1$ fm$^{-2}$
(solid curve). The LUND model \cite{lund} gives a value of 1 fm$^{-2}$. In
Fig.~(\ref{fig1_pt}) we also show the transverse momentum distribution, with
$\alpha$ in Eq.~(\ref{ptd}) set equal to 3.88 (GeV/c)$^{-1}$. Quark
rescattering has very little effect on the transverse momentum produced in the
reaction so that curve is not shown in this figure. We see that a value for
$\Lambda$ near the LUND model value is required by the experimental data. 
Increasing
$\Lambda$ causes the strings to decay faster and hence reduces the
multiplicity. Too low an average multiplicity then propagates through the
other distributions giving results which are low for small rapidity, low for
small fractional longitudinal momentum, and low for all but the smallest
transverse momenta. We have not found a need to fine tune the value of
$\Lambda$; we use the value of 1  fm$^{-2}$ suggested by the LUND model.

\begin{figure}[thbp]
%\vspace*{+2mm}
\begin{minipage}[t]{7.5cm}
{\epsfxsize=7cm\epsfysize=4.7cm \epsfbox{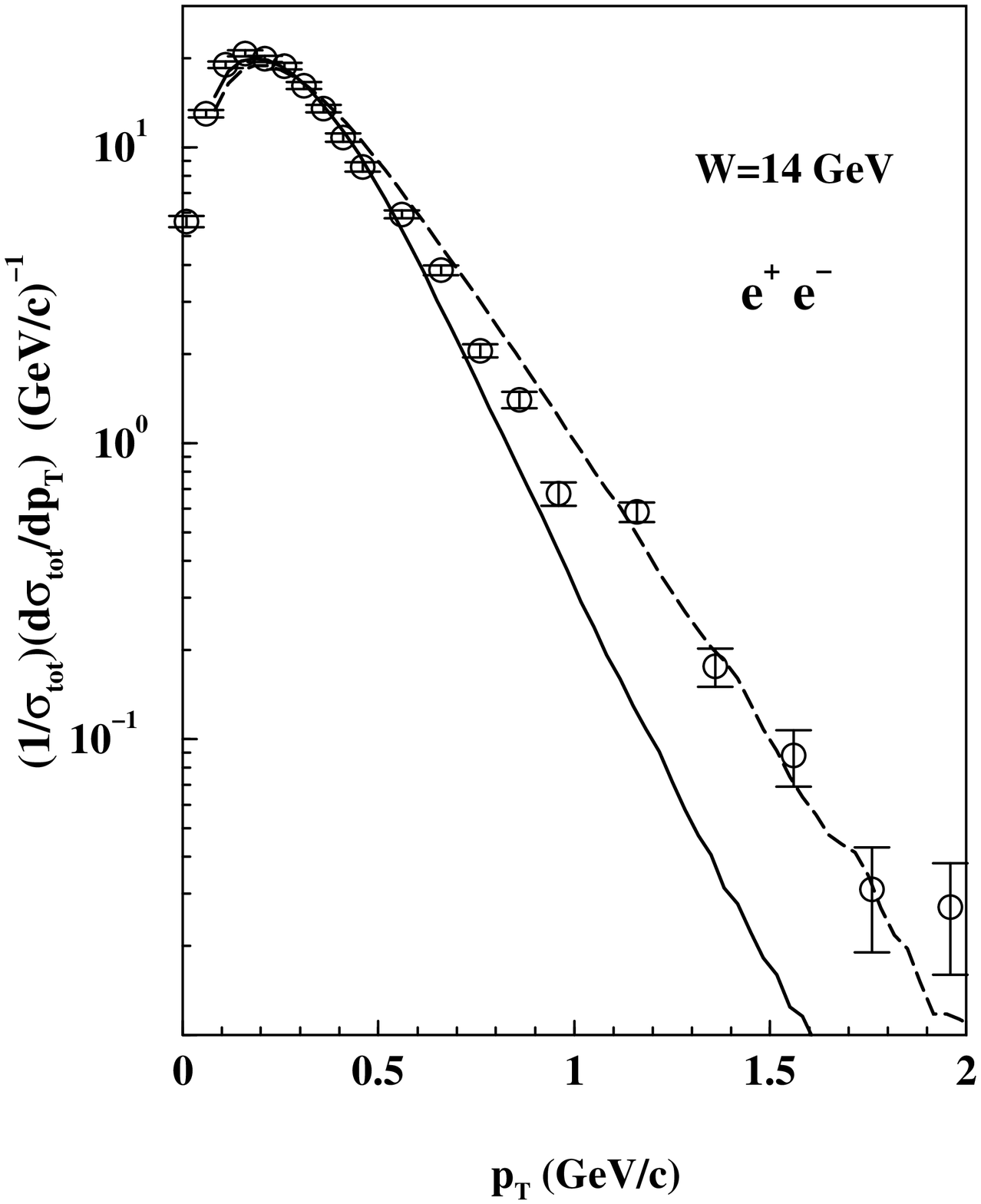}}
\caption[]{\footnotesize  Transverse momentum distributions in  $e^+e^-$ collisions at
$\sqrt{s}=$ 14 GeV. The data are from the TASSO
experiment \protect{\cite{TASSO}}. The solid curve is for $\alpha =$ 3.88 
(GeV/c)$^{-1}$ and the dashed curve is for $\alpha =$ 1.5 (GeV/c)$^{-1}$. }
\label{fig1_trans_a}
\end{minipage} \hfill
\vspace*{-7.1cm}
 \hspace*{6.5cm} \hfill
\begin{minipage}[t]{7.5cm}
{\epsfxsize=7cm\epsfysize=4.7cm \epsfbox{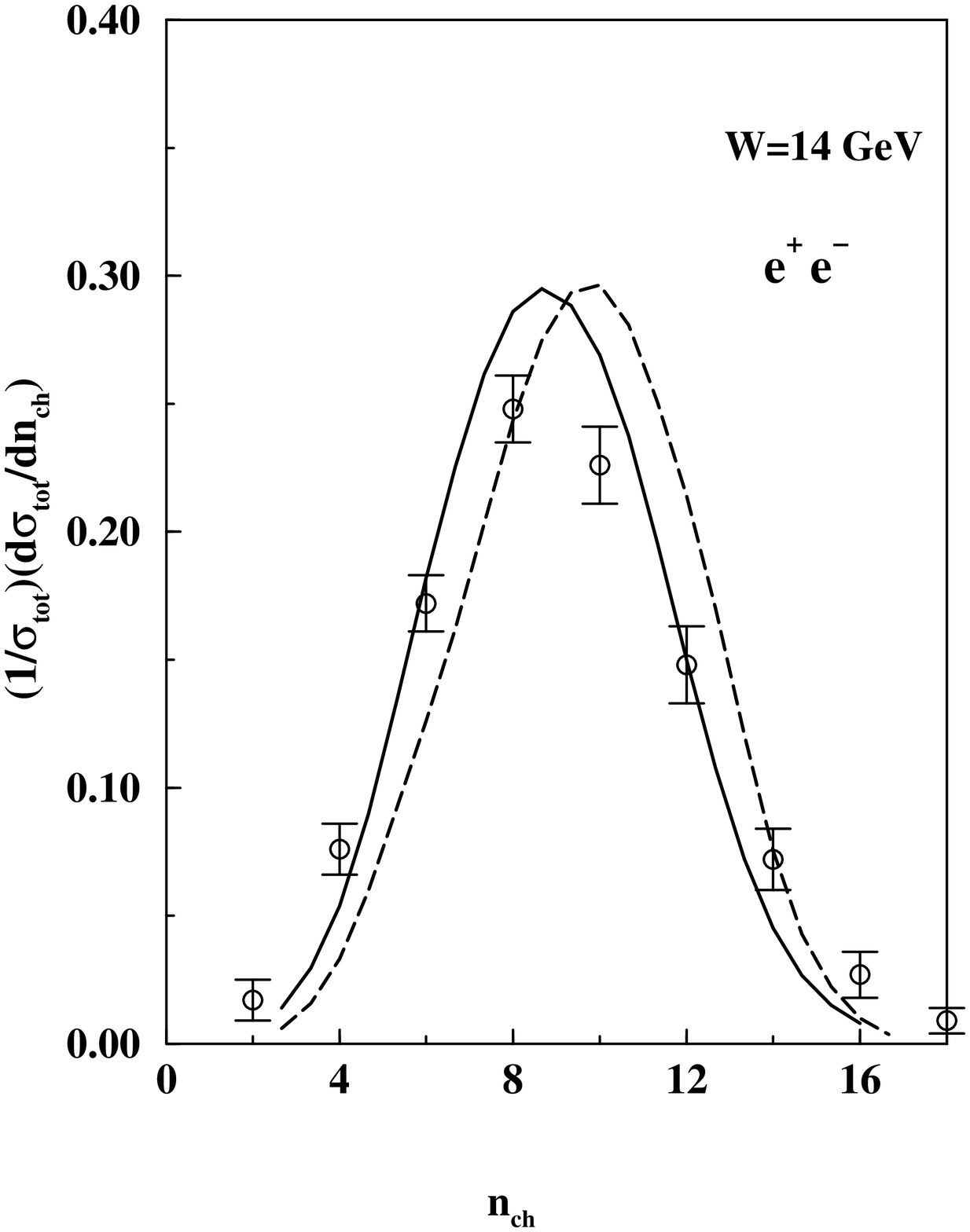}}
%\vspace*{2pt}
\caption[]{\footnotesize  The same as Fig.~({\protect{\ref{fig1_trans_a}}}) except the
multiplicity distribution is given.   }
\label{fig1_multi_a}
\end{minipage}
\end{figure} 
\vspace*{1.5cm}
\begin{minipage}[thbp]{6.5in}
The next parameter to be determined is the parameter $\alpha$ in
Eq.~(\ref{ptd}) --- the parameter which governs the creation of transverse (to
the string) momentum when a string breaks.
This  
\begin{table}[thbp]
\caption{\footnotesize The values of the windows $\delta m_i$ for particles production
identified in the model.  
\label{table1} }
\begin{tabular}{ccccc}
\quad particle\quad & \quad particle mass (GeV) \quad 
& \quad $m_{min}$ (GeV) \quad  & 
\quad $m_{max}$\quad (GeV) & \quad reaction type 
 \quad  \\
\hline
 $\pi$           & 0.14    & 0.14     & 0.42 & $e^+e^-$, $pp$  \\
 $K$             & 0.49    & 0.49     & 0.63 & $e^+e^-$, $pp$  \\
 $p$             & 0.94    & 0.94     & 1.30 & $e^+e^-$, $pp$  \\
 $\eta$          & 0.54    & 0.63     & 0.68 & $e^+e^-$         \\
 $\rho$,$\omega$ & 0.77    & 0.77     & 0.85 & $e^+e^-$         \\
 $K^*$           & 0.89    & 0.89     & 0.98 & $e^+e^-$         \\
 $\eta\prime$    & 0.96    & 0.98     & 0.99 & $e^+e^-$         \\ 
 $f_0$           & 0.98    & 0.990    & 0.997 & $e^+e^-$         \\ 
\end{tabular}
\end{table}
\end{minipage}
parameter  is clearly to be
determined by the transverse momentum distribution in the $e^+e^-$ reactions.
We depict the normalized transverse momentum distribution
$(1/\sigma_{tot})\,(d\sigma/dp_T)$ in Fig.~(\ref{fig1_trans_a}). The solid
curve is for $\alpha=$ 3.88 (GeV/c)$^{-1}$ and the long-dashed curve is for
$\alpha =$ 1.50 (GeV/c)$^{-1}$.  The data for $p_T>$ 1 GeV/c would prefer the
latter value. However, the model does not include some physics, such as gluon
bremsstrahlung. This missing piece of physics is believed to be the source of
the very large $p_T$ events as can be deduced from a change in the slope of
the curve. Moreover, fitting the high energy tail of this reaction amounts to
letting a very small fraction of the total events determine the value of
$\alpha$. We believe the best approach is to fit, in detail, the peak of the
$p_T$ curve which constitutes the majority of the events. Under predicting the tail
then leaves room to add other large momentum transfer physics which would
increase the small number of very high-momentum particles and reproduce the
tail of the distribution. We thus use the value of 3.88 (GeV/c)$^{-1}$. In
Fig.~(\ref{fig1_multi_a}) we show the multiplicity distribution for the two
values of $\alpha$. Lowering $\alpha$ from 3.88 (GeV/c)$^{-1}$ (solid curve)
to 1.50 (GeV/c)$^{-1}$ has the effect of not only increasing the production of
high momentum particles but also increases the average multiplicity, as would
be expected. The multiplicity distribution further indicates that the
preferred value of $\alpha$ is 3.88 (GeV/c)$^{-1}$. We utilize this value and
also find quite satisfactory results assuming $\alpha$ is independent of the
energy of the string.
As noted in the previous section, we have developed \cite{DM98} an approach to
quantize the final state hadrons. The method consists of defining mass
windows which extend from the physical mass of a meson
$m_i$ to $m_i+\delta m_i$. If, when a string breaks, one of the resulting
segments has a mass which lies in the window, the string breaking point is
adjusted slightly to produce a segment with the physical mass. Strings with
the mass of a physical meson are then not allowed to decay. All other strings
are allowed to continue to decay. Some special considerations must also be
applied if the results would lead to a string that could not decay
into two pions; the details are given in Ref.~\cite{DM98}. For the
calculations we will present here, only the masses of the mesons are
quantized. The baryonic strings are allowed to decay as long as the mass of a
baryon is greater than the mass of the nucleon plus a pion. Strings with a
mass smaller than this are identified as a nucleon and not allowed to decay.
Windows for excited baryons can also be used when we wish to
confront data in which excited baryons  
are to be specifically identified. 
It is important
to realize that not allowing a string to decay does {\it not} mean the string
will appear as a particle in the final state.  
The string will continue to
interact, which will excite or de-excite it, after which it  
would continue
through 
\begin{minipage}[thbp]{6.5in}
\begin{table}
\caption{\footnotesize Hadron multiplicities in $e^+e^-$ annihilation events.
The data are from \protect\cite{TASSO} the TASSO collaboration; the
theoretical results are labeled DSP Model.
\label{table2} }
\begin{tabular}{clcccc}
$\sqrt{s}$ (GeV) & 10  & 10 & 30 & 30 \\
meson & TASSO & DSP Model & TASSO & DSP Model \\
\tableline
$\pi$  & 9.8(0.5)  & 9.33 & 15.9(0.7)  & 16.08 \\
$K$    & 1.81(0.09)& 1.83 & 2.96(0.16) & 2.92 \\
$\eta$ & 0.20(0.04)& 0.28 & 0.61(0.07) & 0.49 \\
$\rho\;,\omega$ & 0.65(0.12) & 0.54 & 0.81(0.08)& 0.90 \\
$K^*$  & 0.56(0.06)& 0.65 & 1.2(0.11) & 1.01 \\
$\eta\prime$ & 0.03(0.01) & 0.049 & 0.26(0.1) &0.08 \\
$f_0$        &0.024(0.006)& 0.032 & 0.11(0.04)& 0.059\\     
\end{tabular}
\end{table}
\end{minipage}
the regular decay process. The final state is dynamically determined
when interactions cease to occur.
The mass windows we find from fitting the $e^+e^-$ data are given in
Table~\ref{table1}. We introduce windows and hence quantize the mass of the 
pion, kaon, $\eta$, $\rho$, $\omega$, $K^*$, $\eta^/$, and $f^0$. However,
when calculating results for hadronic collisions during the rest of this work,
we utilize only the windows for the pion and kaon (and the proton) as the data
we examine identify at most these three particles.

A stringent test of this model is to examine the production of individual
mesons as a function of the energy of the $e^+e^-$ collision. The mass windows
are taken to be parameters that are independent of all other quantities. In
Table~\ref{table2}, we present the experimental results for pion, kaon,
$\eta$, $\rho$, $\omega$, $K^*$, $\eta^/$, and $f^0$ production for energies
$\sqrt{s}=$ 10 and 30 GeV. We see that the energy dependence is quite
reasonably produced by this model.

\section{Quark and gluon interactions and $p\,$--$p$ collisions}

Given the model of the nucleon and the model of hadronization presented in
the previous section, we require only a model for the interaction of hadrons
to be able to calculate hadronic collisions. In this section, we will present
a model of the interaction between quarks and gluons, and describe a
technique for incorporating the gluonic content of the nucleon into the
model. These  new features of the model incorporate more completely 
 QCD phenomenology.
  
We have described the nucleon by a distribution of string solutions with the
endpoints of each of the strings identified with a quark, anti-quark, or
di-quark. However,  evidence \cite{close} indicates that a hadron contains a
significant amount of glue. In particular, the momentum fraction of a hadron
carried by the quarks as measured by deep inelastic scattering is \cite{lai}
only 58\% ($Q=1.6$ GeV) with an estimated error of 2\%. The remaining
momentum, 42\%, we will assume is carried by the gluons. We will use this
58\%-42\% division  in our simulations of proton-proton, proton-nucleus, and
nucleus-nucleus collisions unless otherwise stated. For transverse  energies
$E_T$ of jets  produced in $\bar p\,$--$ p$ collisions in the range $50\;$
GeV$<E_T<450$ GeV, and $x$ in the range of $0.06-0.5$, the relative importance
of the three subprocesses -- quark-quark,  quark-gluon, and gluon-gluon --
shifts continuously from being gluon dominated to quark dominated \cite{lai}.
This is also seen in the recent analysis of inclusive jet production cross
section $d\sigma/dE_T$ \cite{CDF}.

We wish to include in our model the significant gluonic content of the
nucleon and the effect it will have on the hadronic collisions.  In order to
do this, we extend  the  model of the nucleon in terms of classical strings.
In addition to the strings whose endpoints are identified as
quarks, we will also include strings which we will identify as pure glue. The
nucleon structure function is then divided between the contribution from the
quarks and the contribution from the gluons, and is of the form
\begin{equation}  
h(x)=C_qq(x)+C_gG(x)\; ,  
\label{qgstructure}
\end{equation}   
where the nucleon structure function is $h(x)$, its quark contribution 
$q(x)$, and gluon contribution $G(x)$, with probabilities $C_q$ and  $C_g$, 
respectively. A mathematical construction identical to the one given in
Ref.~\cite{DM99} allows us to introduce purely gluonic strings and relate the
distribution of their solutions to the measured gluonic structure function.
For the gluon and quark structure functions, we utilized  results
from the Coordinated Theoretical/Experimental Project on QCD and Tests of the
Standard Model (CTEQ) collaboration\cite{lai}. 
We utilize the ``modified minimal
subtraction'' model (also called $\overline{\rm MS}$ or CTEQ4M model) as it is
described as ``this set gives an excellent fit to all data sets" and ``this set
has the best overall quantitative agreement between the next to leading order
QCD theory and the global data on high-energy scattering." This model
parameterizes  the 
structure functions as, 
\begin{equation} 
q(x)=A_0x^{A_1}(1-x)^{A_2}(1+A_3x^{A_4})\;,
\label{struc_form}
\end{equation} 
with the parameters \cite{lai} given in Table \ref{table3}.  
Next the model of the interaction between hadrons is described. We assume
that hadrons interact via the interaction between the quarks, identified as
the endpoints of the quark strings. In addition, we assume an analogous model
for the interactions that involve gluonic strings. The interaction  takes
place between the endpoints of the gluonic strings that contain a finite
fraction of the energy and momentum in the string. Thus we identify the
endpoints of all strings as point like partons and model the interactions as
parton-parton interactions. 

In the parton model assuming factorization \cite{close},  the total
cross-section for $p\,$--$p$ collisions is given in terms of the parton
structure functions $q(x)$, and the elementary parton cross-sections by
\begin{equation}
\sigma_{pp}=\sum_{\alpha,\beta}\int dx_1dx_2d\Omega q_{\alpha}(x_1)q_{\beta}(x_2)
\frac{d\sigma\left(q_{\alpha} q_{\beta} \rightarrow  q_{\alpha} q_{\beta}\right)}
{d\Omega}\; .
\label{sigma_pp}
\end{equation}
where $C_q\,q(x_i)$ in Eq.~(\ref{qgstructure}) is the sum of the up and down
quark structure functions $q_{\alpha}(x_i)$ and $C_g\,G(x_i)$ is the gluon
structure function. Thus $q_\alpha (x_i)\,dx_i$ is the probability  of
finding a parton of flavor $\alpha$ with momentum fraction $x_i$ in nucleon
$i$,  and $d\sigma/d\Omega$ is the elementary scattering cross-section for
the partons.  If the parton scattering cross-sections are known, then the
total hadronic cross-section can be calculated. 
Our model of the interaction of the partons allows them to interact if the
partons from two different strings come within an interaction radius $r_g$ 
of each other.
Therefore, by considering a large number of $p\,$--$p$
collisions at fixed value of impact parameter $b$ we can estimate
scattering probability 
$P(b)$ as ratio of reactions with interacting partons to the total number
of reactions.
This allows us to compute the total $p\,$--$p$ scattering cross section
in the form
\begin{equation}
\sigma_{pp}=2\pi\int_{0}^{\infty}P(b)bdb\; .
\label{P(b)}
\end{equation}

\begin{figure}[thbp]
\begin{minipage}[t]{7.5cm}
{\epsfxsize=7cm\epsfysize=4.7cm \epsfbox{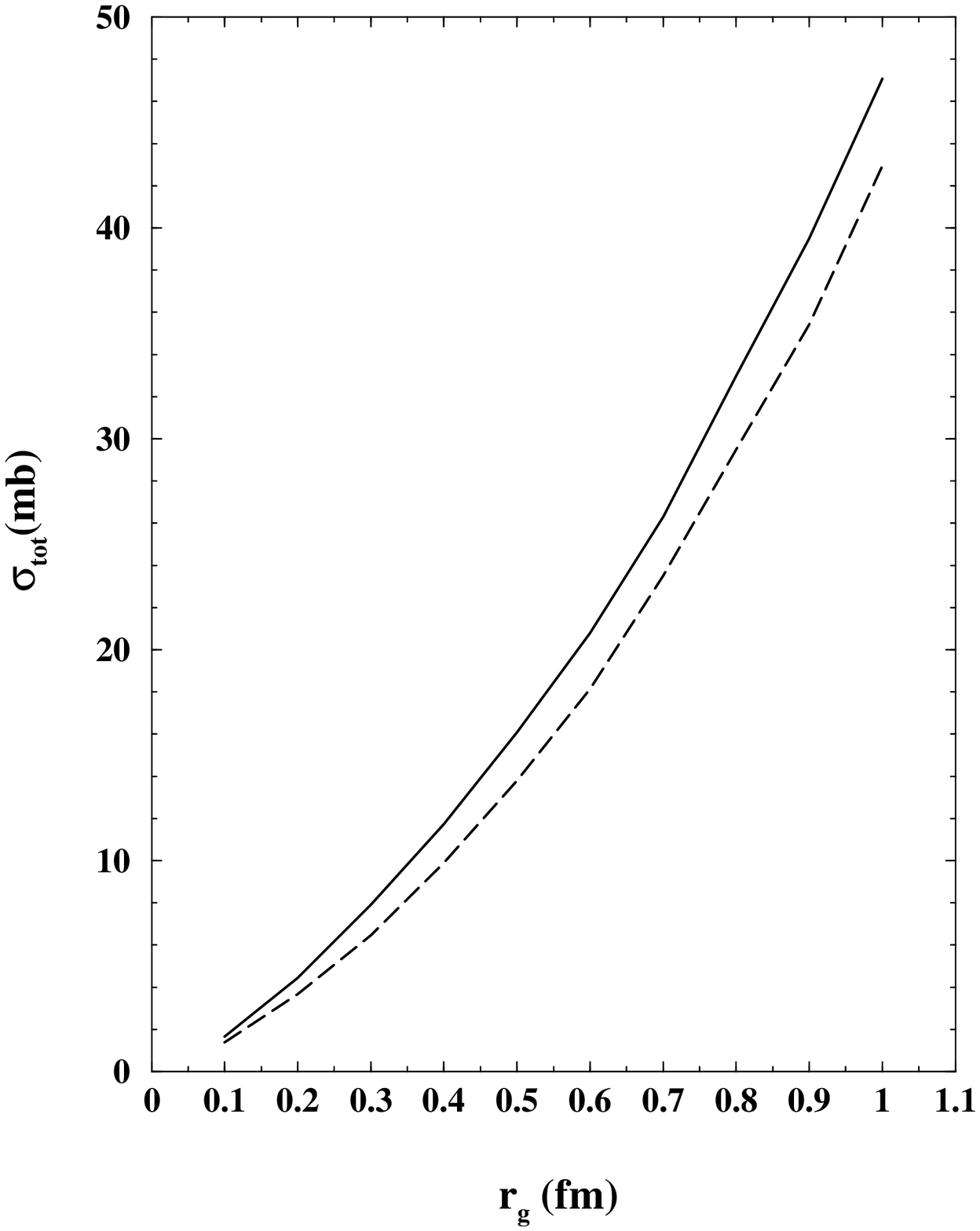}}
\caption[]{\footnotesize The total  \protect$p\,$--$p$ cross-section  as a function of $r_g$, the model
parton-parton interaction radius, at $\sqrt s=30$GeV. The solid curve assumes
the proton is completely made up of quark strings, while the dashed curve
assumes the proton is completely made up gluonic strings. The difference
results from the different structure functions, as can be seen in
Eq.~(\protect{\ref{sigma_pp}}).}
\label{fig1_sigma}
\end{minipage} \hfill
\vspace*{-8.6cm}
\hspace*{6.5cm}\hfill
\begin{minipage}[t]{7.5cm}
{\epsfxsize=7.1cm\epsfysize=4.7cm \epsfbox{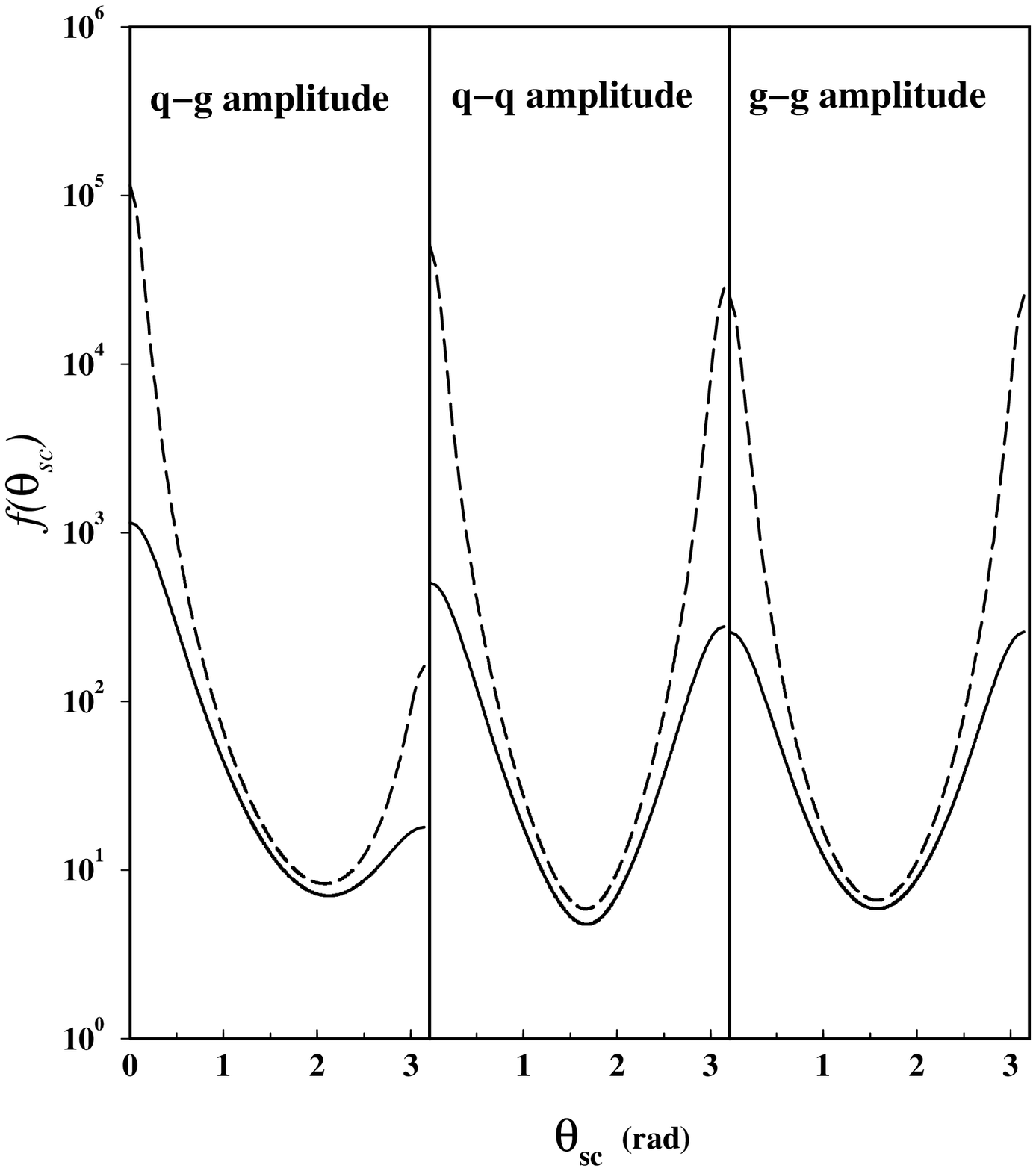}}
\caption[]{\protect\footnotesize Perturbative QCD predictions for the 
differential scattering cross sections for parton-parton (quark-gluon, 
quark-quark,
gluon-gluon) scattering as a function of the center-of-mass scattering
angle $\theta_{sc}$ in radians.The solid curves are for a center-of-mass
energy  $\sqrt{\hat s}=1$ GeV while the dashed curves are for
$\sqrt{\hat s}=10$ GeV.}
\label{fig2_amplit}
\end{minipage}
\end{figure}

Results of these calculations are shown in Fig.~\ref{fig1_sigma} 
for proton reactions simulated by quark strings and gluonic strings.
Thus an interaction radius can be related to the 
 ~\cite{DMthesis,atomki} 
%\vspace*{0.5cm}
\begin{minipage}[thbp]{6.5in}
\begin{table}[thbp]
\caption
{\footnotesize Parameters in Eq.~(\protect{\ref{struc_form}}) for the CTEQ4M parton 
structure functions measured at $Q_0=1.6$ GeV as taken from
Ref.\protect\cite{lai}.} 
\label{table3}
\begin{tabular}{p{1.6cm}p{1.6cm}p{1.6cm}p{1.6cm}p{1.6cm}p{1.6cm}p{3.0cm}}
Parton & $\; A_0$ & $\; A_1$  & $\; A_2$ &  $\; A_3$ & $\; A_4$  & 
\% Momentum \\
\hline
$xd_v$ & 0.640 & 0.501  & 4.247  & 2.690 & 0.333  &\  \ \ 11.2 \\
$xu_v$ & 1.344 & 0.501  &3.689   & 6.402   & 0.873 &\  \ \ 30.6  \\ 
$xg$   & 1.123 & -0.206 &4.673   & 4.269  & 1.508  &\  \ \ 41.7 \\
\end{tabular}
\end{table} 
\end{minipage}
\hspace*{-7mm}  experimentally measured $p\,$--$p$ total
inelastic scattering cross section.
  An experimental total inelastic cross
section of 28 mb~\cite{breakstone}, gives a parton interaction radius 
of $r_g=$ 0.7 fm as pictured in Fig.~\ref{fig1_sigma}. The solid curve is 
the result calculated assuming the proton is totally made up of quark strings
while the dashed curve is the results assuming only gluonic strings. The
difference arises from using the different structure functions, as would be
expected from 
Eq.~(\ref{sigma_pp}). We see that, although there is a difference between the
curves, it is not large. This allows us to assume a single interaction radius
for all of the parton-parton interactions, which, unless otherwise stated,
will be taken as 0.7 fm.

As is conventional, we divided parton-parton scattering into two
contributions, a soft part and a hard part. By definition, the hard part is
taken directly from perturbative QCD  calculations. We utilize the
theoretical parton-parton cross-sections as given in
Refs.~\cite{peskin,cutler}:
\begin{equation}
\frac{d\sigma}{d\hat t}\left(q_a q_b \rightarrow q_a q_b\right) = 
\frac{4\pi\alpha_s^2}{9\hat s^2}
\left[\frac{\hat s^2+\hat u^2}{(\hat t-m_g^2)^2}+\delta_{ab}\left(\frac
{\hat t^2+\hat s^2}{(\hat u-m_g^2)^2}-\frac{2}{3}\frac
{\hat s^2}{(\hat u-m_g^2)(\hat t-m_g^2)}\right)\right],\\
\label{qq}
\end{equation} 
\begin{equation}
\frac{d\sigma}{d\hat t}\left(qg \rightarrow qg\right) = 
\frac{4\pi\alpha_s^2}{9\hat s^2}
\left[-\frac{\hat u}{(\hat s-m_g^2)}-\frac{\hat s}{(\hat u-m_g^2)}
+\frac{9}{4}\left(\frac{\hat s^2+\hat u^2}{(\hat t-m_g^2)^2}\right)\right],\\
\label{qg}
\end{equation}
\begin{equation}
\frac{d\sigma}{d\hat t}\left(gg \rightarrow  gg\right) = 
\frac{9\pi\alpha_s^2}{2\hat s^2} 
\left[3-\frac{\hat t \hat u}{(\hat s-m_g^2)^2}-\frac{\hat s \hat u}
{(\hat t-m_g^2)^2}-\frac{\hat s \hat t}{(\hat u-m_g^2)^2}\right],
\label{gg}
\end{equation}
where $\alpha_s^2=g_s^2/(4\pi)$ and $m_g$ is the gluon mass.  The Mandelstam
variables $\hat s$, $\hat t$ and $\hat u$ for the massless quarks are given
by $\hat s=2p_i\cdot p_j$, $\hat t=-2p_i\cdot p_1=-\hat s(1-\cos\theta)/2$,
and $\hat u=-2p_i\cdot p_2=-\hat s(1+\cos\theta)/2$.  These cross sections 
describe the scattering of two partons with incoming momentum states $p_i$
and $p_j$, and outgoing momentum states $p_1$ and $p_2$, and $\theta$ is the
angle between $p_i$ and $p_1$ as measured in the center-of-momentum frame  of
the partons.  We choose a gluon mass $m_g=0.25$ GeV which gives a range of
the interaction consistent with the 28 mb $p\,$--$p$ inelastic cross-section. 

For the hard interactions, we utilize the appropriate parton-parton 
scattering cross-sections given in Eqs.~(\ref{qq}-\ref{gg}) and pictured in
Fig.~\ref{fig2_amplit}. If there is sufficiently large momentum transfer,
then, as in deep inelastic scattering, the elementary interaction  takes place
very rapidly  compared to the internal time scale of the  hadronic 
interaction. In this case, the lowest order QCD predictions should describe
the process.  It is seen in Fig.~\ref{fig2_amplit} that these cross sections
are forward and backward peaked. This forward and backward peaking is not
present in proton-proton scattering, giving strong evidence that the momentum
transfers are not generally large enough for the perturbative calculations to
be valid. 

At smaller momentum transfers, one would expect the interaction to be poorly
modeled by perturbative QCD as the coupling constant $\alpha$ becomes large
\cite{peskin} for small momenta, something seen in lattice QCD
calculations \cite{digiacomo}. We might expect the interaction to be
dominated by a large number of non-perturbative soft gluon exchanges.
Assuming incoherence in the  collision process, the total momentum transfer
would be a sum of many elementary  momentum exchanges among the quanta of the
color field, $Q=\sum_jk_j$. The transverse components of the $k_j$'s are  in
general rather small and pointing in random directions. Similar to a random
walk in the transverse plane,  $Q$ would be expected to obey a Gaussian
distribution with the average $\langle Q\rangle$ on the order of the 
hadronic mass scale. However, the Gaussian distribution falls off too rapidly
in comparison to data, so in our model we assumed for soft momentum transfers
an exponential distribution, 
\begin{equation}
 f(Q)\,Q\,dQ\propto \exp ({-\chi Q})\,Q\,dQ, 
\label{soft} 
\end{equation}        
with the constant $\chi$ of the order of an inverse
hadronic mass scale. We utilize this single cross section for all the
parton-parton soft interactions. Unfortunately it is very difficult to
distinguish hadronic jets initiated by gluons from those initiated by quarks.
Thus this minimalist assumption of equality of soft scatterings for the
parton-parton interactions will not contradict any data and will be used
unless or until we find a reason to assume otherwise.

We divide the parton-parton scattering into hard and soft events by
introducing a threshold parameter $\sqrt{\hat s_{min}}$, where $\sqrt{\hat
s}$ is the center-of-mass energy of the two partons. If the center-of-mass
energy of the interacting partons is greater than $\sqrt{\hat s_{min}}$ then
the partons interact via the hard interactions, Eqs.~(\ref{qq}-\ref{gg}), and
if the center-of-mass energy is less than $\sqrt{\hat s_{min}}$ they interact
via the soft interaction, Eq.~(\ref{soft}).

To implement the hard and soft scattering cross sections numerically, we must
scatter two partons whenever they come within a radius $r_g$. The scattering
must obey the probability distributions implied by the differential cross
sections given in the above formulae. For the hard scattering, we generate a
random scattering angle chosen with a weight given by the appropriate cross
section in Eqs.~(\ref{qq}-\ref{gg}) and then numerically scatter the partons
through the appropriate angle. For the soft scattering, we generate a random
momentum transfer with a weight given by Eq.~(\ref{soft}), calculate the
corresponding scattering angle, and numerically scatter the partons through
this angle. 
Proton-proton scattering is then  simulated on the computer by
generating a proton as a collection of strings chosen  to reproduce the
quark and gluon structure functions. A second proton is similarly generated.
They are separated and boosted toward each other  along trajectories with an
impact parameter $b$. We evolve the strings 
according to their equations of
motion and model the interactions of the partons and the hadronization process
as described above. Just as is done in an experiment, we count the
number of final state particles, their energy and momentum, and construct the
distributions which we wish to compare to experiment. The process is repeated
by varying the impact parameter and by generating additional models of the
proton, all constrained to provide on the average the exact experimental
structure functions.
\begin{figure}[thbp]
%\vspace*{-10mm}
\begin{minipage}[t]{7.5cm}
{\epsfxsize=7cm\epsfysize=4.7cm \epsfbox{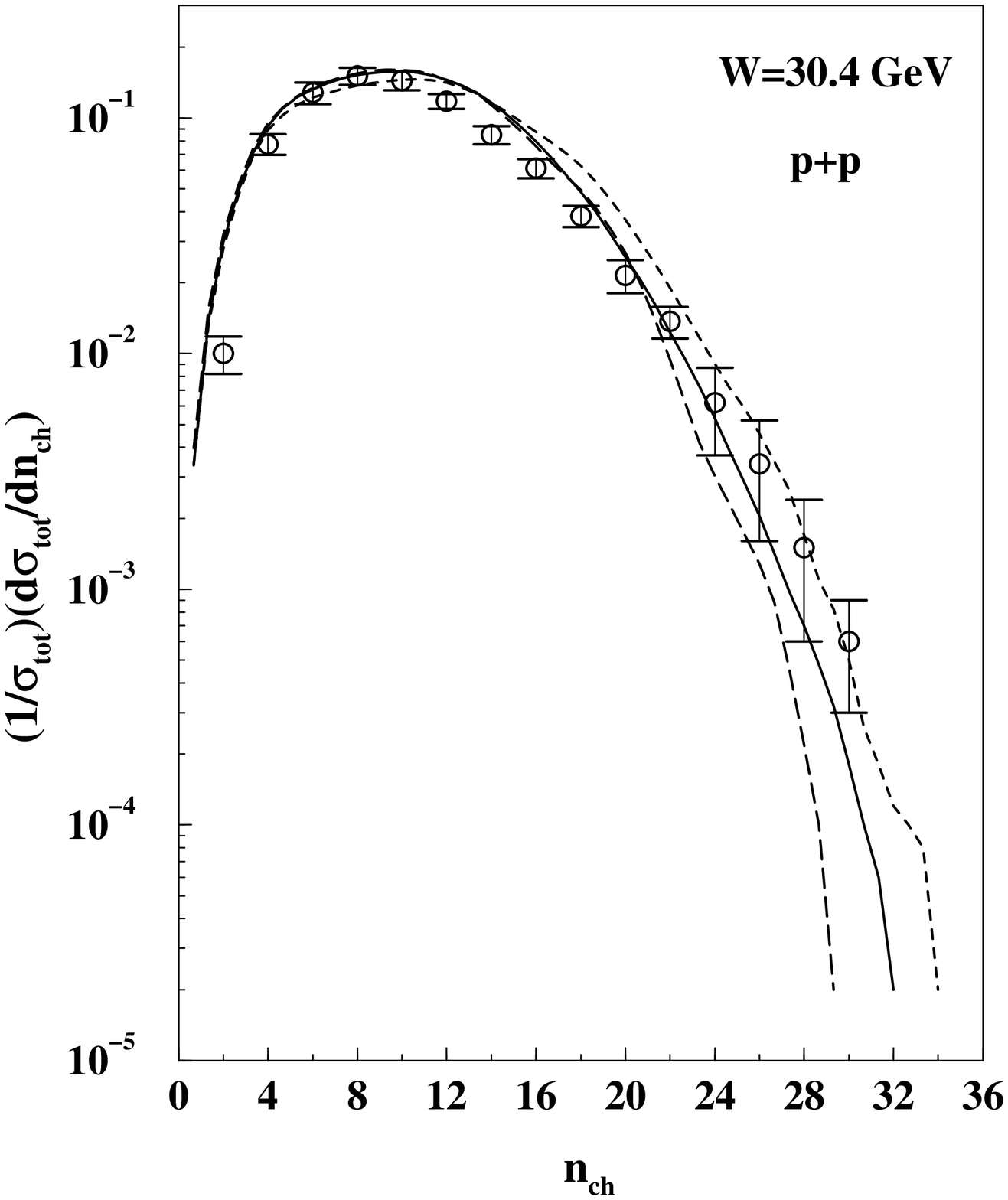}}
\caption[]{\footnotesize Multiplicity distribution for $p\,$--$p$ collisions at $\sqrt{s}=$30.4 GeV. The
data are from \protect\cite{breakstone}. The long-dashed curve is the result
assuming no hard parton-parton scattering, the solid line is the theoretical
result with approximately 4\% hard scattering and the remainder
soft scattering according. The short-dashed curve results from a calculation 
in which approximately 19\% is hard scattering. }
\label{pp_multi1}
\end{minipage} \hfill
\vspace*{-9.2cm}
 \hspace*{6.5cm} \hfill
\begin{minipage}[t]{7.5cm}
{\epsfxsize=7cm\epsfysize=4.7cm \epsfbox{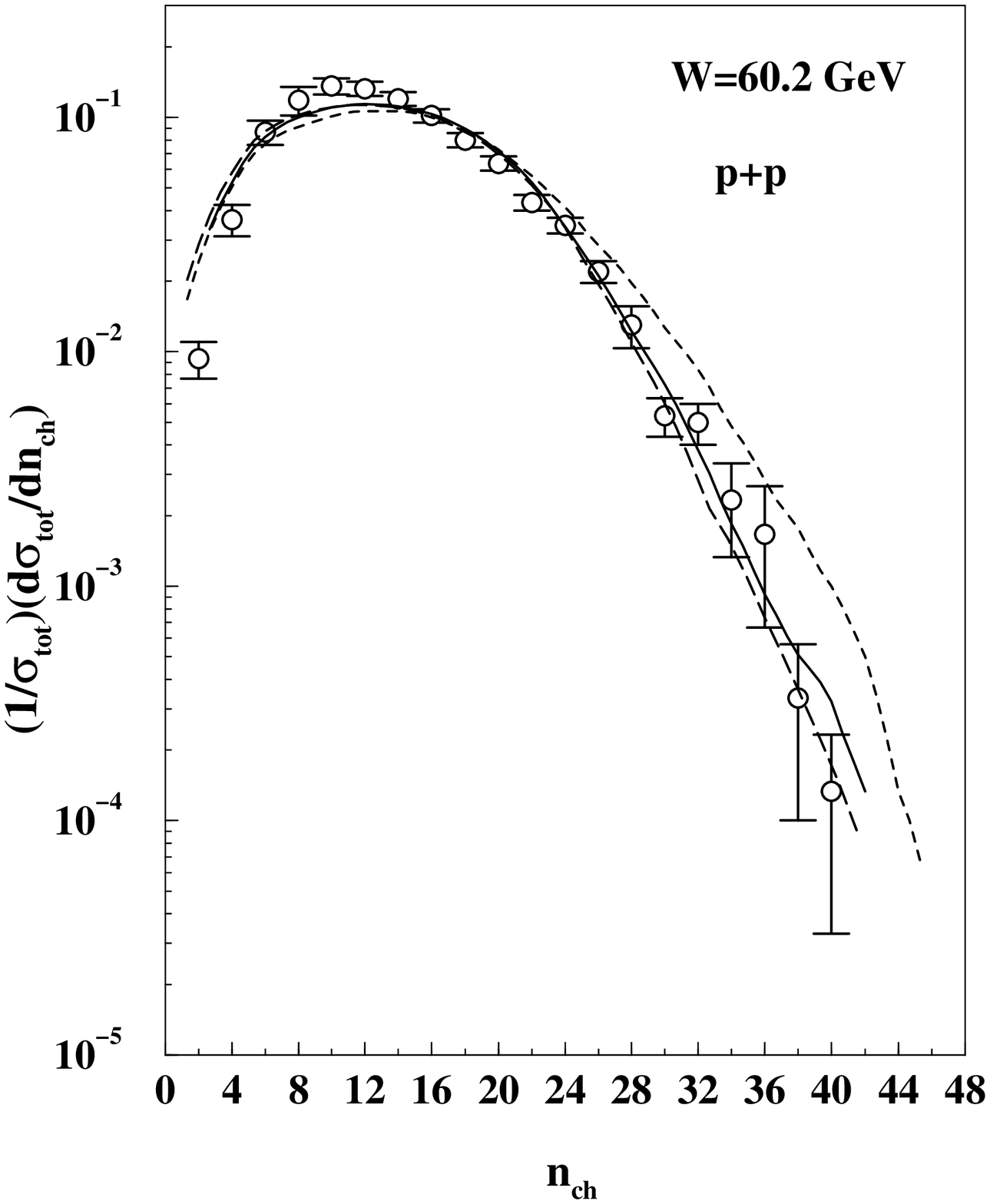}}
%\vspace*{2pt}
\caption[]{\footnotesize The same as Fig.~\ref{pp_multi1} except $\sqrt{s}=$60.2 GeV   }
\label{pp_multi2}
\end{minipage}
\end{figure} 
\vspace*{3.5cm}

From $p\,$--$p$ scattering we must determine the remaining parameters of the
model. Intuitively, one would expect an increase in high multiplicity events
with an increase in the amount of hard scattering. We find this to be true as
depicted in in Figs.~\ref{pp_multi1} and \ref{pp_multi2} where we show the
experimental multiplicity \cite{breakstone} for $p\,$--$p$ collisions at 
$\sqrt{s}=$ 30.4 and 60.2 GeV. We use protons that are composed of 42\%
gluonic strings for these calculations. We also fix the soft momentum transfer
parameter $\chi$ in Eq.~(\ref{soft}) to be 0.7 fm$^{-1}$. The long-dashed
curves in both figures do not include any hard scattering. They fall
rather close to the data indicating that the fraction of hard scattering is
small. The short-dashed curve uses $\sqrt{\hat s_{min}}= 2$ GeV. The percent
hard scattering is then 18\% at 30.4 GeV and 24\% at 60.2 GeV. These curves
produce too many high multiplicity events. We conclude that  $\sqrt{\hat
s_{min}}=$4.0 GeV which produces the solid curves.  This corresponds to 4\% at
30.4 GeV and 5\% at 60.2 GeV of the scatterings being hard and reproduces
quite well the high multiplicity tail. We conclude that hard scattering is not
negligible but that it occurs only a small percent of the time.

\begin{figure}[thbp]
%\vspace*{-10mm}
\begin{minipage}[t]{7.5cm}
{\epsfxsize=7cm\epsfysize=4.7cm \epsfbox{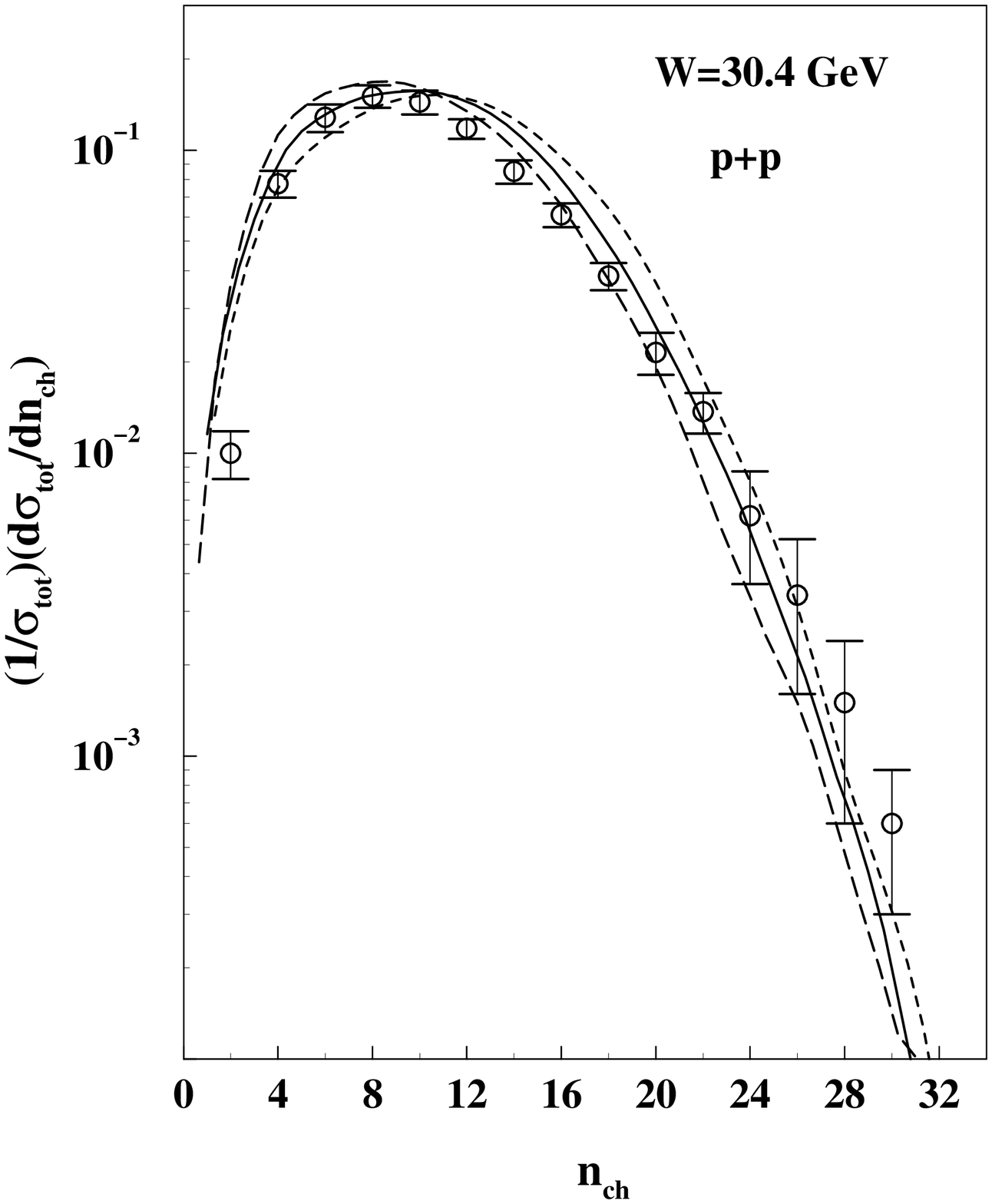}}
\caption[]{\footnotesize  Multiplicity distribution for $p\,$--$p$ collisions at $\sqrt{s}=$30.4
GeV. The data are from \protect\cite{breakstone}. The long-dashed curve is a
calculation with $\chi=1.0$; the solid line is $\chi=0.7$, the
value we choose for the model; and the short-dashed curve is the result
with $\chi=0.4$ }
\label{pp_multi3}
\end{minipage} \hfill
\vspace*{-7.7cm}
 \hspace*{6.5cm} \hfill
\begin{minipage}[t]{7.5cm}
{\epsfxsize=7cm\epsfysize=4.7cm \epsfbox{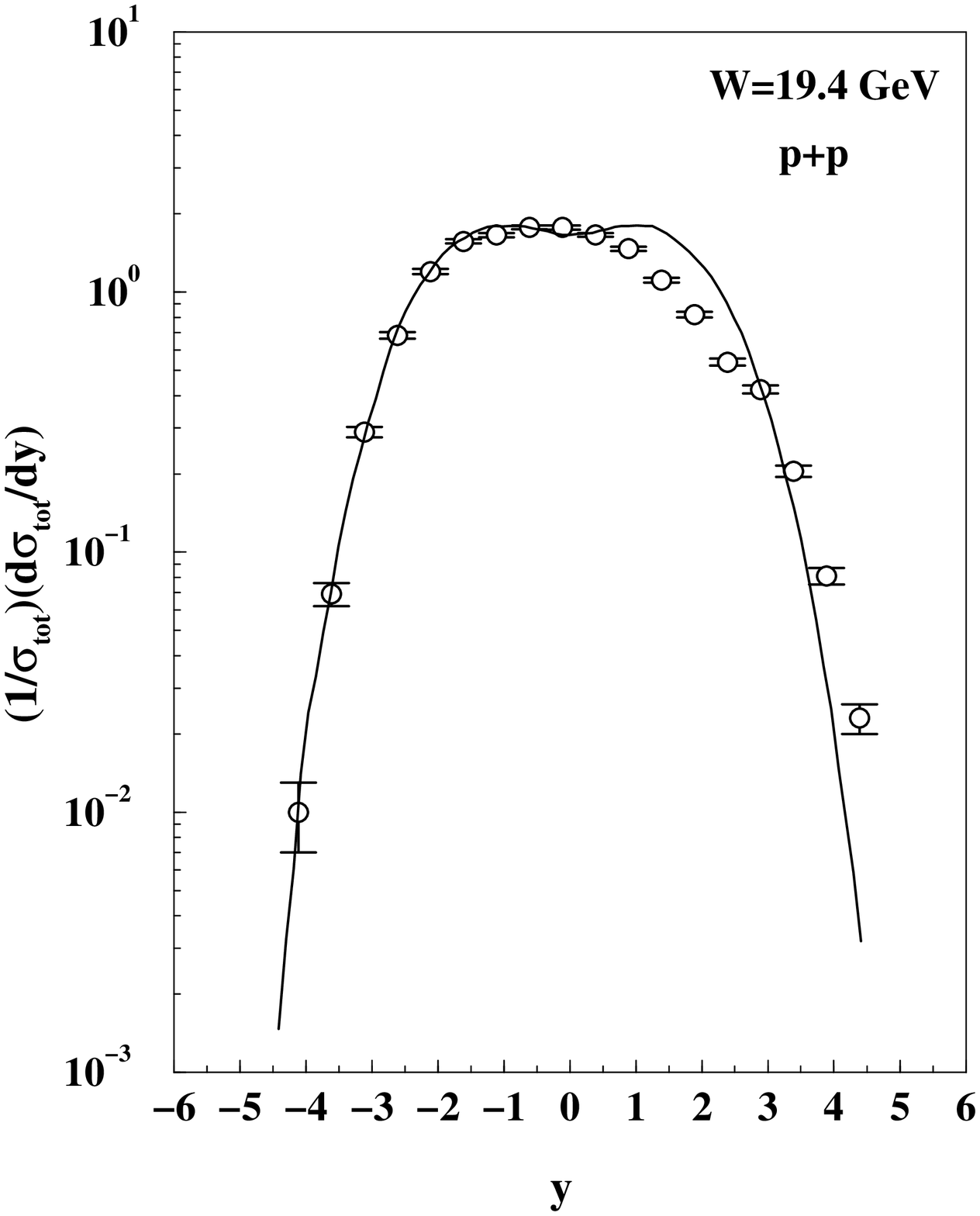}}
%\vspace*{2pt}
\caption[]{\footnotesize  Rapidity distribution for $p\,$--$p$ collisions at 
$\sqrt{s}=$19.4. GeV. The data are from \protect\cite{NA35} and the solid curve
is the prediction of the DSP model developed here.
 The limited identification of
protons leads to a distortion of the experimental rapidity distribution
in the target hemisphere.  }
\label{pp_rapid}
\end{minipage}
\end{figure}

We also need to determine the momentum transfer parameter $\chi$ that governs
the soft interaction, Eq.~(\ref{soft}). In Fig.~(\ref{pp_multi3}) the
multiplicity distribution for $p\,$--$p$ scattering at $\sqrt{s}=$ 30.4 GeV is
shown. The three curves represent three values for $\chi$ --- $\chi=$ 0.4,
0.7, and 1.0 fm$^{-1}$. The inverse value of $\chi$ corresponds to the average
soft momentum transfer during the parton scattering. A decrease in the value
of $\chi$ causes an increase in  the inelasticity  of the process and
therefore an increase in the multiplicity of the reaction, as can be seen in 
Fig.~(\ref{pp_multi3}). We find that $\chi=$ 0.7 fm$^{-1}$ gives the best fit,
and utilize that value for our
%\vspace*{0.5cm}
\hspace*{-8mm}
model.  It is interesting to compare
Fig.~(\ref{pp_multi1}) with Fig.~(\ref{pp_multi3}).  The momentum transfer
parameter $\chi$ affects mostly the distribution near the peak multiplicity,
while the amount of hard scattering affects mostly the tail of the
distribution. 
We may also look at the experimental $p_T$ distributions of the final state 
particles. The $p_T$ distributions are to a good approximation exponential, at
least over the range $p_T=0.1$ to $1.1$ GeV/c \cite{CERN}. We present in
Table~\ref{table4} the slope parameters extracted from the data \cite{ros75}
and those calculated from our model. We take the fraction of the proton that
is composed of gluon strings \cite{close} to be 42\% as extracted from data.
However, it is interesting to ask whether this affects the calculated
$p\,$--$p$ results. We also present in Table~\ref{table4} the slope
parameters extracted from $p_T$ distributions calculated using protons
composed totally of quark strings  and totally of gluon strings. 
We see   that  mixed state of 58\% quark and 42\% gluon content in the proton
gives a better fit to data   and indicates that significant fraction of the
strings should be gluonic.
 
For completeness we also present in Fig.~(\ref{pp_rapid}) the measured
rapidity distribution for $p\,$--$p$ collisions at $\sqrt{s}=$ 19.4 GeV. The
data are from Ref.~\cite{NA5}. The calculation utilizes our final choice for
all of the parameters.  These are given in Table~\ref{table5} with a
description of how they were determined. These results are completely 
predictive as this data was not used in determining any of the parameters. 
Notice the
excellent results for negative rapidity. Also note 
that the experimental
results are not symmetric as they must be. 
\begin{minipage}[thbp]{6.5in}
\begin{table}[thbp]
\caption
{\footnotesize Slope parameters for the $p_T$ distributions of the final state pions, kaons
and protons in $p\,$--$p$ collisions at $\sqrt{s}=$ 30.4 GeV. The experimental
values are from Ref.~\protect\cite{ros75}. The results of calculations using
the DSP model are presented, each using a different fraction, as labeled, of
quarks versus gluons in the composition of the nucleon.
}
\label{table4}
\begin{tabular}{p{1.6cm}p{3.3cm}p{3.5cm}p{3.3cm}p{3.3cm}}
Particle &  Slope ($(GeV/c)^{-1}$)& Slope ($(GeV/c)^{-1}$)&Slope ($(GeV/c)^{-1}$)& Slope 
($(GeV/c)^{-1}$)\\
         &  Experiment             & 58\%quark/42\%gluon        
	 &  100\% quark &  100\%gluon       \\
\hline	   
Pions      &  \quad 6.5-7.0      &\quad 6.70         & \quad 6.34   
   &\quad 7.58   \\ 
Kaons      &  \quad 5.0          &\quad 4.97         & \quad 4.62   
   & \quad 6.46   \\      
Protons    &  \quad 4.0	         &\quad 4.12         & \quad 3.70   
   &  \quad 5.79   \\   
\end{tabular}
\end{table}  
\end{minipage}
This is a fixed target experiment and
 it is noted that for the positive rapidity region there was less than
complete separation between pions and protons. We are thus very satisfied
 with
the excellent results we find for negative rapidity, and discrepancies are
largely a result of experimental distortion~\cite{NA5}. 
We describe the model as minimalist
because the number of parameters is quite small in comparison to other models.
In addition, once these parameters have been determined from $e^+e^-$ and
$p\,$--$p$ collisions, the model is completely predictive.  In the next section
we describe how we construct a nucleus out of a distribution of nucleons. A
$p\,$--$A$ or $A$--$A$ collision is then modeled by constructing the
appropriate initial state, evolving it in time on the computer according to
our covariant and dynamical model, and then collecting statistics on the final
state particles, just as do the experimentalists. 
 
\section{Proton-Nucleus Collisions} 

Proton-nucleus reactions are a definitive test of
the string-parton model. The parameters of the  model are completely determined from
$e^+e^-$ and $p\,$--$p$ data. The model is now predictive and thus comparison to
$p\,$--$A$ and $A$--$A$ data are a validation of the assumptions of the model, if the
model compares well with the data.
 
Although the main focus of relativistic heavy-ion physics is on $A$--$A$ collisions and
the formation of the quark-gluon plasma, $p\,$--$A$ provide an opportunity to study the
dynamics of the hadronic interaction in a situation that is simpler than an $A$--$A$
collision. As QCD is not capable of confronting this data, models are necessary. 
In order to create the densities and temperatures to form the plasma, a nucleon in one
nucleus must collide multiple times with other nucleons 
and deposit a large amount of
energy in the interaction region. The evidence of the occurrence of multiple
collisions comes directly from experimental investigations \cite{barton}. 
The $p\,$--$A$
system provides insight into how these multiple collisions come about.

In the $p\,$--$A$ system at high-energies the identity of the proton and 
the particles
created from its excitation is, to a great extent, maintained throughout 
the collision.
This is indicated experimentally by distributions having a peak tied 
to the kinematics of
the incident proton. The model and the experimental data then provide 
insight into the
importance of multiple collisions for the incident proton and on how 
the second and
subsequent collisions are modified \cite{fritiof,rqmd} by having 
undergone a previous
collision. This modification is significant because a head-on 
collision of two protons
can create a highly excited string which will then
\begin{minipage}[thbp]{6.5in}
\begin{table}
\caption
{\footnotesize Parameters in the DSP model, their value, and how they are determined. In
addition the mass windows used to quantize the mesonic strings are given
in Table I.}
\label{table5}[thbp]
\begin{tabular}{p{1.4cm}p{6.1cm}p{4.4cm}p{2.6cm}}
Symbol     & Description                      & Method of determination        & Value\\  
\hline 
$\kappa$   & String tension             & Regge slope                    
   & $0.88$ GeV/fm\\ 
$\Lambda$  & String decay rate          &$e^+e^-$ multiplicity data      
   & $1.0$ fm$^{-2}$\\
$\alpha$   & $p_T$ production constant  &$e^+e^-$ $p_T$ data             
   & $3.88$ (GeV/c)$^{-1}$\\
$r_g$      & parton interaction radius         & $pp$ total cross-section       
   & $0.7$ fm          \\
$\chi$     & soft-momentum transfer constant  & $pp$ and $e^+e^-$  data  
   & $0.7$ (GeV/c)$^{-1}$\\
$\sqrt{\hat s_{min}}$ &energy threshold for hard scattering  
   & $pp$ multiplicity data & $4$ GeV \\
\end{tabular}
\end{table}
\end{minipage} 
collide again 
before it has had time
to decay. This phenomenon is included quite naturally in the DSP 
model since the
interactions occur between quarks and the strings are evolved 
and decay in space-time
according to the Lagrangian and the invariant decay scheme. 
The ability to vary the
target is then very useful as this variation changes 
the distance through the nuclear
medium that the incident proton must travel. 
The second peak in the distribution occurs
at kinematics that indicate an origin in the 
$A$ nucleons of the ion. For this kinematic
region, there are $A$ particles with nearly 
equal momentum, i.e. within a Fermi momentum
of each other, and thus multiple scatterings will 
be more significant. This still remains
a simpler situation than occurs in an $A$--$A$ system 
and thus can provide additional
insights into the dynamics. Other theoretical models 
do not seem to have published
results for the $p\,$--$A$ data so we are not able 
to compare our results to theirs.

There is one last theoretical piece yet needed 
to model $p\,$--$A$ collisions. We need
to generate a model for a nucleus containing $A$ 
nucleons. We employ a Fermi-density 
distribution 
\begin{equation} 
\rho(r)=\frac{\rho_0}{1+\exp[(r-c)/a_0]}\;,
\end{equation} 
with  parameters $a_0=0.55$ fm the thickness of the nuclear surface, 
$c=1.18A^{1/3}-0.48$ fm the half-density radius, and $\rho_0\simeq 0.07$ fm$^{-3}$ the
central density\cite{bergmann}. The nuclear volume is then populated with nucleons, each
constructed as a distribution of strings as described earlier, with the probability
density $\rho(r)$ determining the distribution of the nucleons. However, two nucleons in
the same nucleus are not allowed to be within an interaction range. If the Monte Carlo
chooses a location such that the quarks that make up a proton would be interacting with
other quarks, another position is generated until a location is found such that the
nucleus will not be composed of particles scattering from each other and causing the
nucleus to react with itself before the physical collision begins. The nucleon is then
offset from the $z$-axis by an impact parameter $b$ which is either fixed or distributed
over a range with the weight  $2\pi bdb$ and the proton and the nucleus are translated in
the longitudinal direction such that they do not overlap. The nucleus and the proton are
then boosted towards each other with the collision Lorentz factor $\gamma$ to reach the
energy of the reaction, and the time evolution proceeds via  the string equations of
motion.

\begin{figure}[thbp]
%\vspace*{-10mm}
\begin{minipage}[t]{7.5cm}
{\epsfxsize=7cm\epsfysize=4.7cm \epsfbox{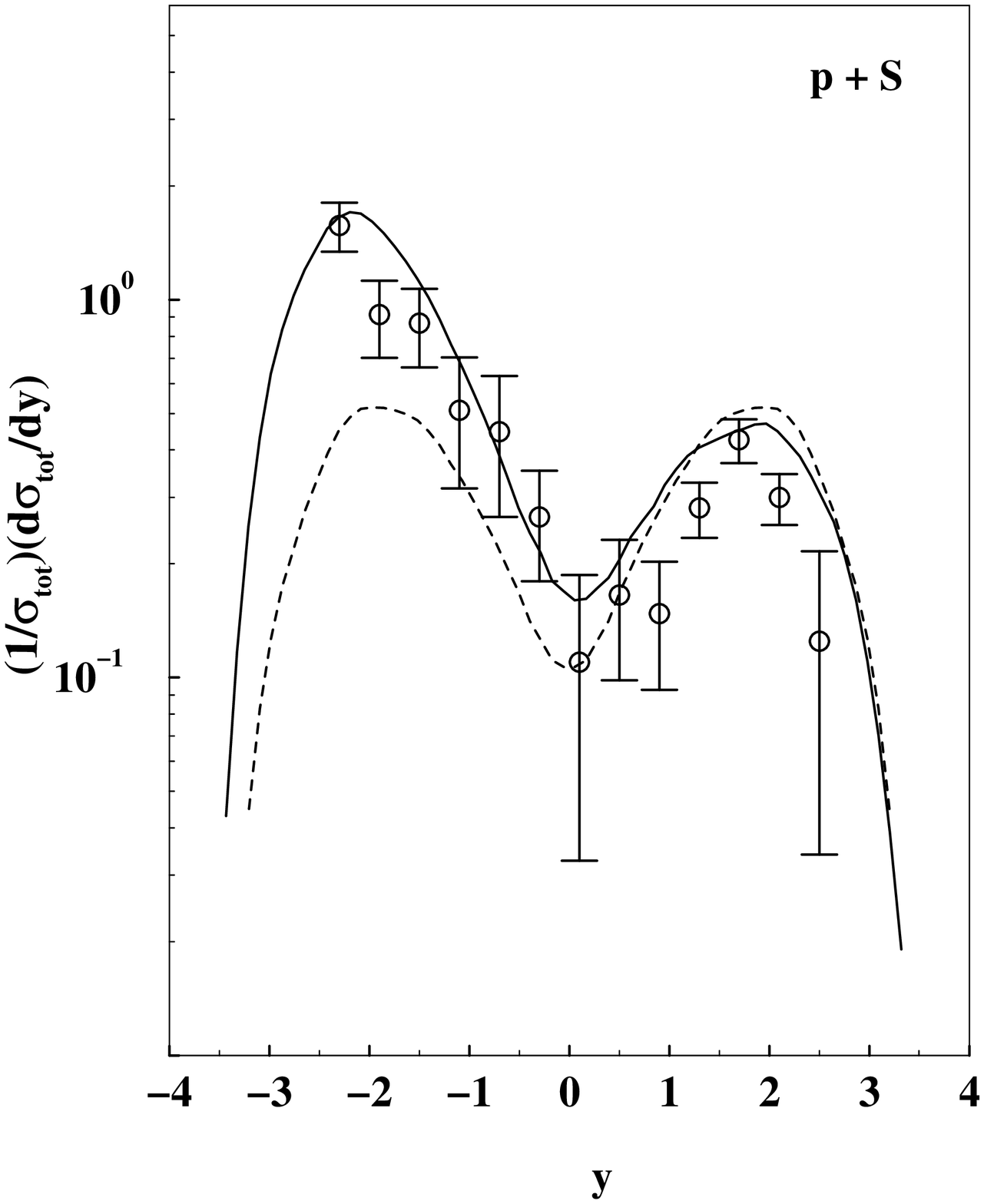}}
\caption[]{\footnotesize Rapidity distribution of net nucleons for minimum bias $p$+$S$ collisions at 
200 GeV per nucleon. The data are from \protect\cite{NA35} and the solid curve
is the prediction of the DSP model developed here. The dashed curve is 
the prediction of the DSP model for $p\,$--$p$ collisions.  }
\label{pS_fig1}
\end{minipage} \hfill
\vspace*{-7.7cm}
 \hspace*{6.5cm} \hfill
\begin{minipage}[t]{7.5cm}
{\epsfxsize=7cm\epsfysize=4.7cm \epsfbox{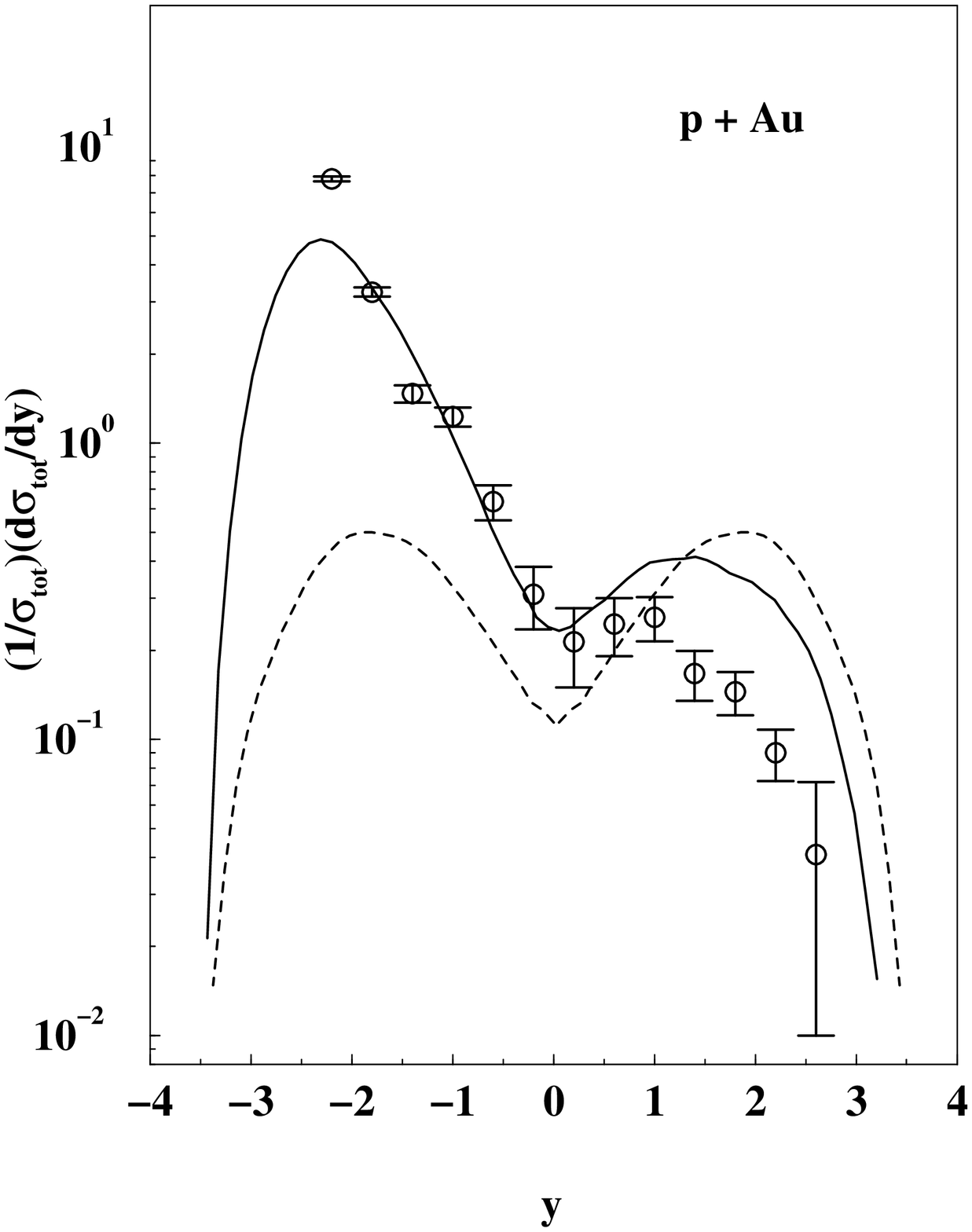}}
%\vspace*{2pt}
\caption[]{\footnotesize  The same as \protect\ref{pS_fig1} 
 except the reaction is $p$+$Au$.  }
\label{pAu_fig1}
\end{minipage}
\end{figure}

We here apply the DSP model to minimum bias data \cite{NA35} for $p$+$S$ and $p$+$Au$
collisions  at 200 GeV per nucleon. The rapidity distributions of net nucleons for
minimum bias $p$+$S$ and $p$+$Au$ \cite{comment} are shown in Figs.~\ref{pS_fig1} and
~\ref{pAu_fig1}.
The data are for net protons while our model does not distinguish
between the proton and the neutron. 
To account for this difference we multiply the data
in the target hemisphere, $y < 0$, 
by two for isoscalar(equal number of neutrons
and protons) nuclei($p$+$S$) and by a factor
of $A/Z$ for other cases. This result in a factor of $2.5$
for $p$+$Au$ reaction. For the projectile hemisphere the data and the
calculation are for one nucleon (here a proton) so this data is not multiplied by two.
Thus the  plots are not net protons, but net nucleons. The solid curves in these figures
are the predictions of the DSP model.

Near midrapidity, $y=0$, for both reactions the model and the data agree quite well. For
the projectile hemisphere,  $y > 0$, the results for sulfur, Fig.~\ref{pS_fig1}, show a
possible and small discrepancy with the data as the largest two rapidity points lie
slightly below the theory.
\vspace*{2.0cm}
\hspace*{-7mm}
The dashed curve in these figures represents the results of
our model for $p\,$--$p$ collisions. There is very little difference between the
$p\,$--$p$ collision curve and the $p\,$--$A$ curve in this hemisphere. This indicates
that for this kinematic region only a single proton scattering has occurred. As a check
on our computer program, we restricted the calculation to a single nucleon scattering and
found that we did indeed recover the $p\,$--$p$ results pictured. If there is a
discrepancy for the largest rapidity, its origin must be in the model of the $p\,$--$p$
collision. For the gold target, the situation in the projectile hemisphere is very
different. The data no longer show a distinct peak. This is a clear indication that their
have been multiple collisions and significant stopping of the proton. The theory shows a
broadened peak shifted to lower rapidity as compared to the  $p\,$--$p$ results (the
dashed curve) indicating the existence of multiple scatterings. However, the theory is
not producing as much stopping as is seen in the data. On the other hand, looking at the
target hemisphere, $y < 0$, the data and the theory are quite compatible. Some caution
must be made in drawing conclusions that make use of the  lowest ($y\leq -2.8$) and
highest ($y\geq 2.8$) rapidity points These data \cite{NA35} were extrapolated on the
basis of  a thermal fit. This was necessary  because at low transverse momenta a cut was
applied at $0.3$ GeV. Also in comparing rapidity distributions for different systems one
has to keep in mind that different forward energy triggers may affect the spectra 
differently \cite{NA35}. In summary, we  find satisfactory results except for the
projectile hemisphere for the gold data, where the theory appears to underpredict the
stopping of the incident proton. 
 
\begin{figure}[thbp]
%\vspace*{-10mm}
\begin{minipage}[t]{7.5cm}
{\epsfxsize=7cm\epsfysize=4.7cm \epsfbox{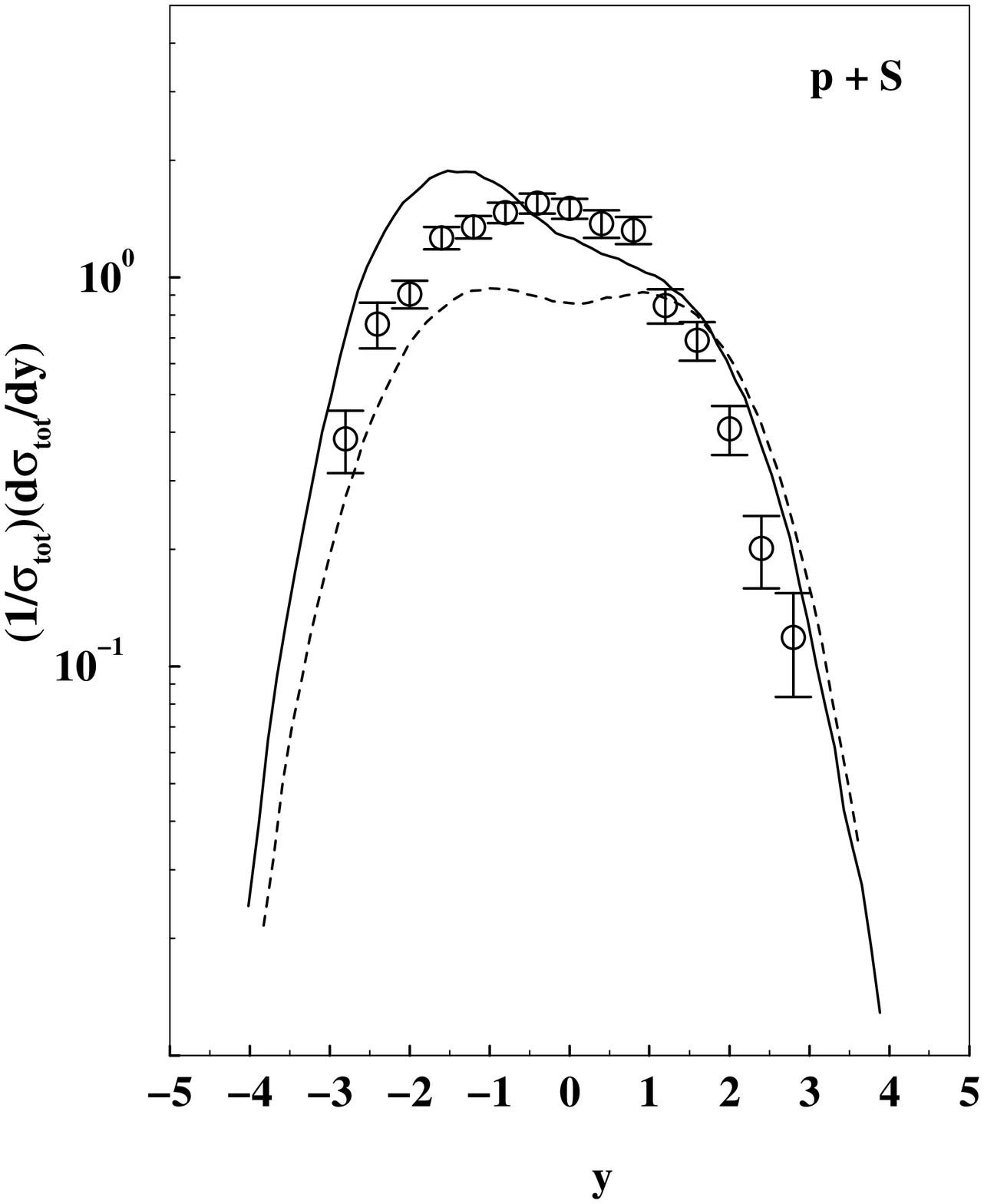}}
\caption[]{\footnotesize Rapidity distribution of negatively charged hadrons
for minimum bias $p$+$S$ collisions at 
200 GeV per nucleon. The data are from \protect\cite{NA35} and the solid curve
is the prediction of the DSP model developed here. The dashed curve is 
the prediction of the DSP model for $p\,$--$p$ collisions.  }
\label{pS_fig2}
\end{minipage} \hfill
\vspace*{-7.7cm}
 \hspace*{6.5cm} \hfill
\begin{minipage}[t]{7.5cm}
{\epsfxsize=7cm\epsfysize=4.7cm \epsfbox{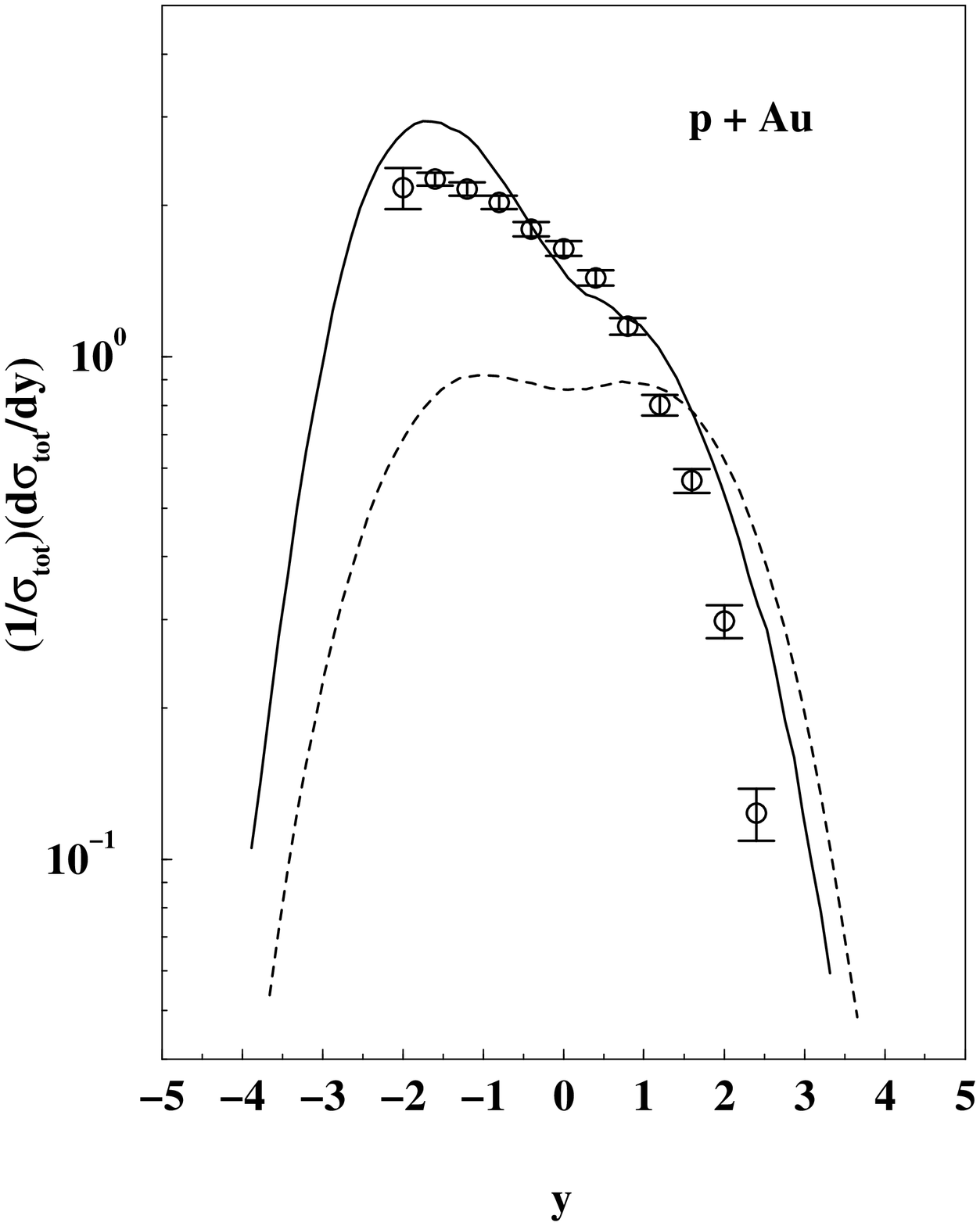}}
%\vspace*{2pt}
\caption[]{\footnotesize  The same as \protect\ref{pS_fig2} 
 except the reaction is $p$+$Au$.  }
\label{pAu_fig2}
\end{minipage}
\end{figure}

\begin{figure}[thbp]
%\vspace*{-10mm}
\begin{minipage}[t]{7.5cm}
{\epsfxsize=7cm\epsfysize=4.7cm \epsfbox{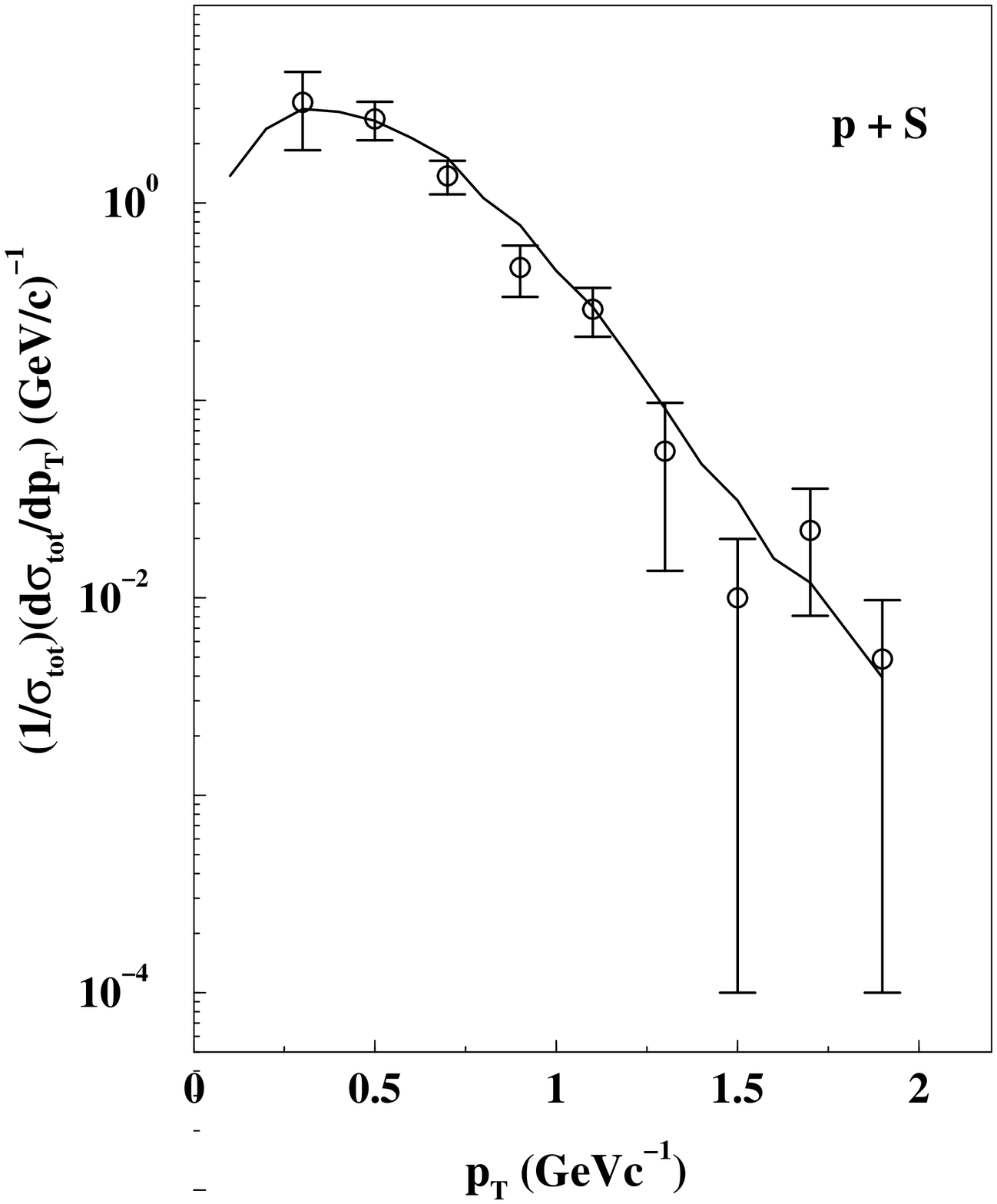}}
\caption[]{\footnotesize Transverse momentum distributions of net nucleons 
for minimum bias  $p+S$ collisions at 
200 GeV per nucleon. The data are from \protect\cite{NA35} and the solid curve
is the prediction of the DSP model developed here. }
\label{pS_pT}
\end{minipage} \hfill
\vspace*{-7.2cm}
 \hspace*{6.5cm} \hfill
\begin{minipage}[t]{7.5cm}
{\epsfxsize=7cm\epsfysize=4.7cm \epsfbox{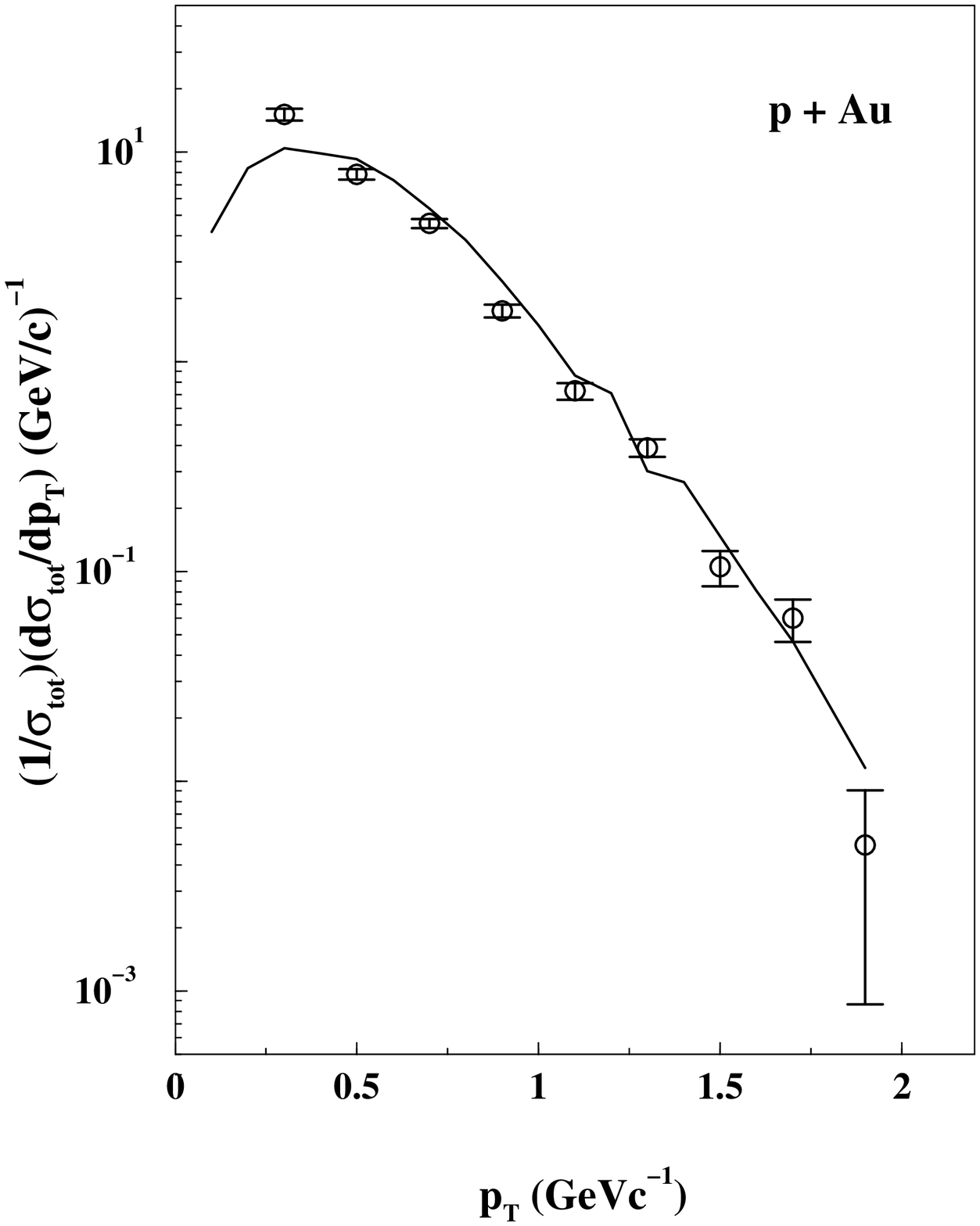}}
%\vspace*{2pt}
\caption[]{\footnotesize  Transverse momentum distributions of net protons 
for minimum bias  $p+Au$ collisions at 
200 GeV per nucleon. The data are from \protect\cite{NA35} and the solid curve
is the prediction of the DSP model developed here.   }
\label{pAu_pT}
\end{minipage}
\end{figure}

The rapidity distributions of negatively charged hadrons produced in minimum bias $p$+$S$
and  $p$+$Au$ collisions are shown in Figs.~\ref{pS_fig2} and ~\ref{pAu_fig2}. The dashed
curve is again the results of the DSP model for a nucleon-nucleon collision. In both the
theory and the data, the multiplicity of negative hadrons increases with target mass and
the maximum shifts to lower rapidities, indicating \cite{NA35} $\pi^-$ production after
multiple collisions. 
We also see that the high rapidity $y\ge 1.7$ region for $p\,$+$S$
collisions is very near the $p\,$+$p$ result indicating that this region is determined by
a single scattering. Although the theory produces the very large enhancement for
$p\,$--$Au$ near $y=-2$ over the $p\,$--$p$ curve, the theory has a tendency to preserve
a slight peak near the target rapidity that is not seen in the data.
\vspace*{2.0cm}
 The degree of thermalization in heavy-ion collisions and  the possible effects of
collective flow on the spectra can be studied by examining the transverse degrees of
freedom. We calculate the spectra of protons, pions and kaons  and compare to data from
experiments NA35 and NA44 from CERN \cite{NA35,NA44}. Figs.~\ref{pS_pT} and ~\ref{pAu_pT}
show the transverse momentum distribution  of net nucleons (scaled by a factor of two to
represent nucleons rather than protons) for $p+S$ and $p+Au$ collisions at 200
GeV$\cdot$A together with calculations of the DSP model. The calculations agree quite 
 well
with data.
\vspace*{0.5cm}
\hspace*{-7mm}
 
An inverse slope parameter (temperature) has been extracted from the data and
can be calculated in our model. The results for protons, kaons, and pions are given in
Table \ref{table6}. We utilized data \cite{NA44} for $K^-$ in the midrapidity region
$k_T=\sqrt{p_T^2+m_K^2}-m_K\leq0.84$ GeVc$^{-1}$ and for pions in the midrapidity region
$k_T=\sqrt{p_T^2+m_{\pi}^2}-m_{\pi}\leq0.64$ GeVc$^{-1}$. Very reasonable agreement is
found for these parameters. In addition the pion spectra generated within the model are
not exponential for all values of  $m_T$. They  show an increase with increasing $p_T$
which is also true of experimental data and other theoretical predictions
\cite{NA44,cronin,helios}.
We have found that the results of the model for $p\,$--$A$ collisions are often
quantitatively in agreement with the data, but there exists in the net nucleon rapidity
distribution for $p\,$--$Au$ an indication of more stopping of the incident proton in the
data than is found in the theory. The rapidity distributions for negative hadrons would
indicate a small increase in the hadronic interactions that would broaden the
distribution of the produced hadrons. Other theoretical models have not published results
for $p\,$--$A$ collisions so we cannot say how general these results might be. In
comparison with the $p\,$--$p$ data, all of the general features found in the experiments
in going to $p\,$--$A$ are contained in the theory. This gives us confidence that
the model is capturing much of the important underlying physical phenomena.

\section{Nucleus-Nucleus Collisions}
The spectra of produced particles and participant nucleons measured over a broad acceptance
in rapidity $y$ and transverse momentum $p_T$ address many questions important to an
understanding of the overall reaction dynamics. It is only in $A$--$A$ collisions that one 
would expect to create sufficient energy densities to create the QGP. Furthermore, in 
$p\,$--$A$ collisions it is difficult to disentangle the effects of rescattering of the 
spectator matter from  the more interesting participant matter dynamics. In central $A$--$A$ 
collisions, the importance of rescattering among spectator matter is minimized and the 
dynamics of of the highly excited, very dense, created matter becomes dominant.

\begin{figure}[thbp]
%\vspace*{-10mm}
\begin{minipage}[t]{7.5cm}
{\epsfxsize=7cm\epsfysize=4.7cm \epsfbox{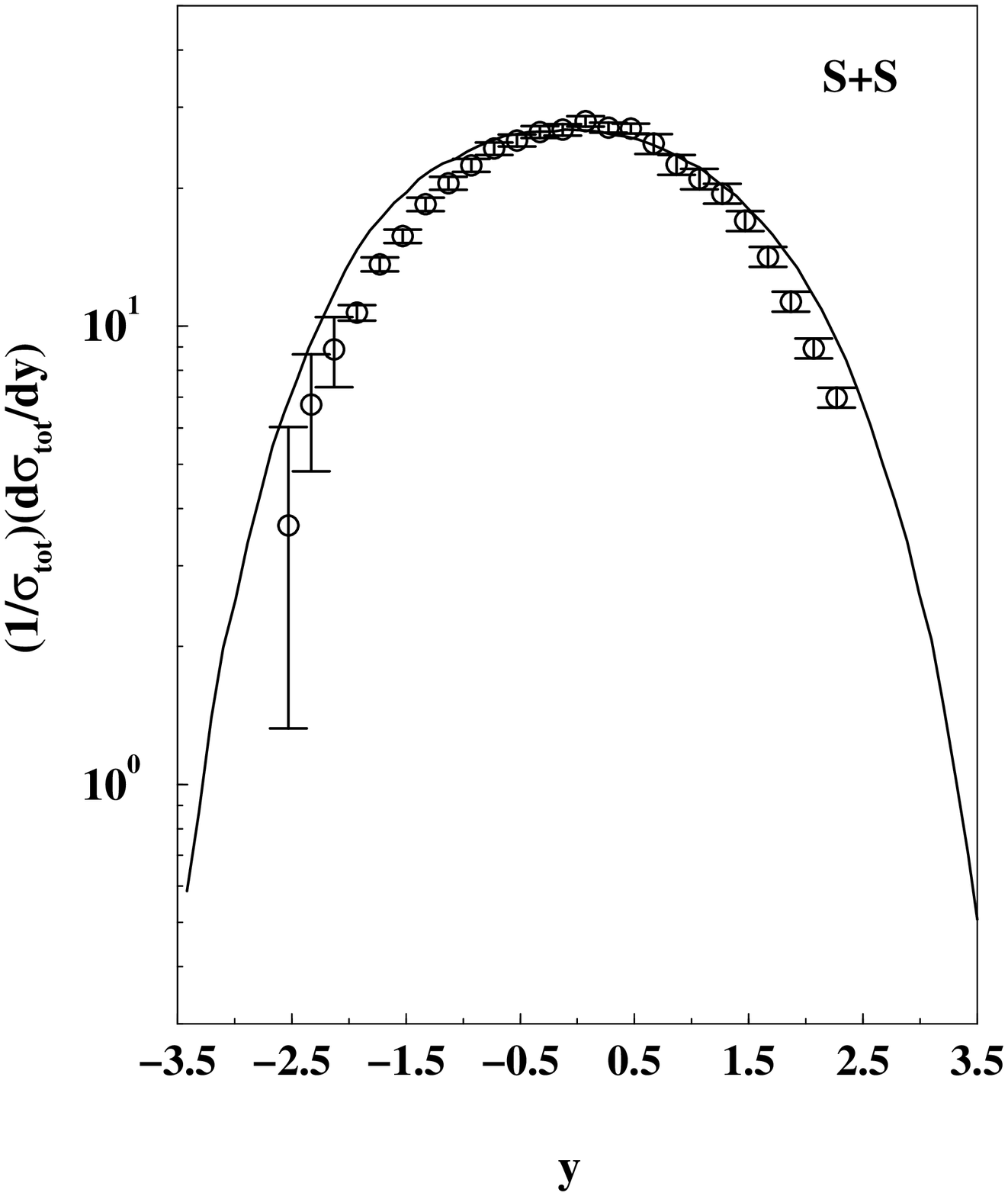}}
\caption[]{\footnotesize Rapidity distribution of negatively charged hadrons
for central $S$+$S$ collisions at 
200 GeV per nucleon. The data are from \protect\cite{NA35,NA35_SS} and the solid curve
is the prediction of the DSP model developed here.  }
\label{fig_SS_neg_rap}
\end{minipage} \hfill
\vspace*{-7.2cm}
 \hspace*{6.5cm} \hfill
\begin{minipage}[t]{7.5cm}
{\epsfxsize=7cm\epsfysize=4.7cm \epsfbox{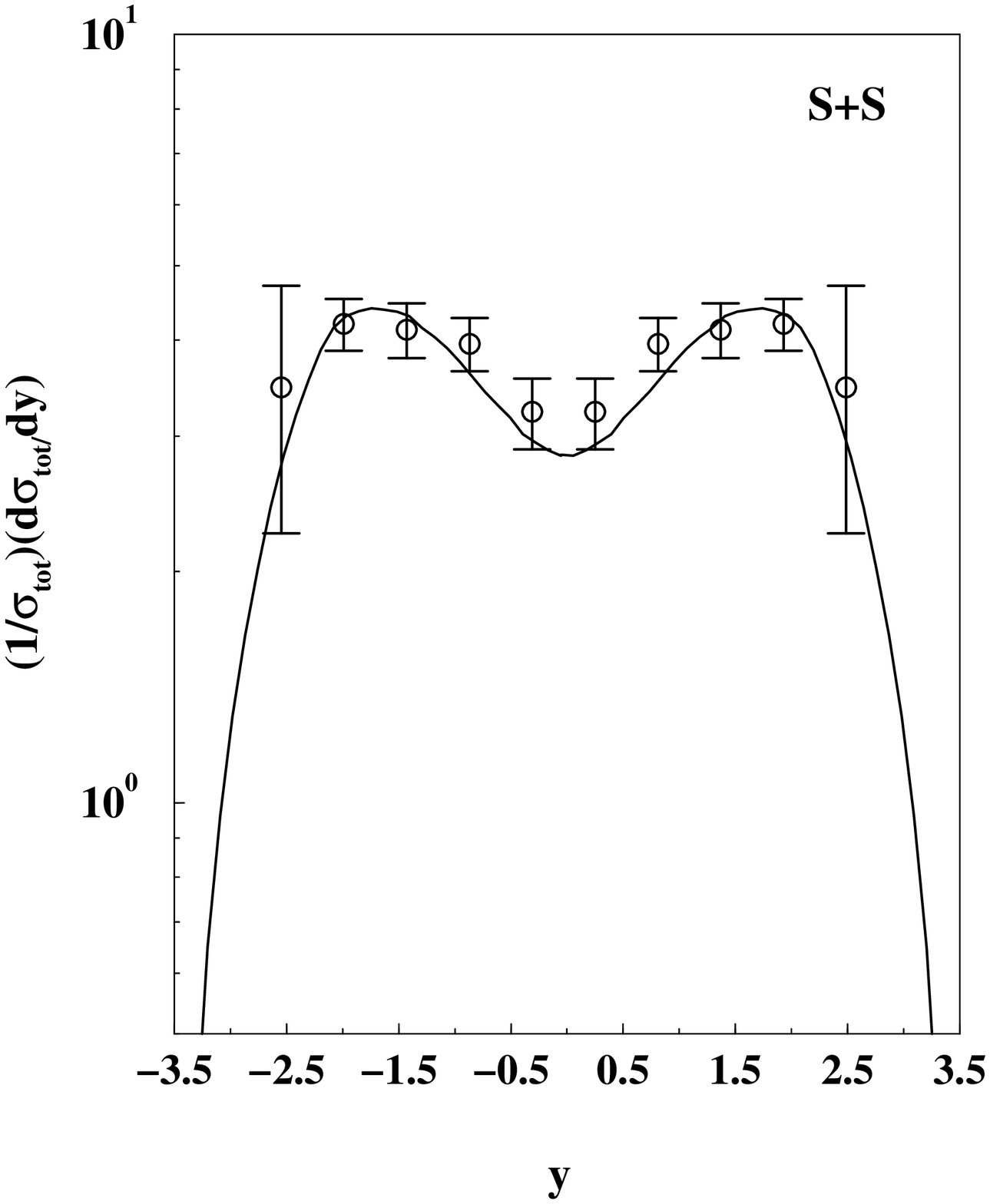}}
%\vspace*{2pt}
\caption[]{\footnotesize  Rapidity distribution of  net protons
for central $S$+$S$ collisions at 
200 GeV per nucleon. The data are from \protect\cite{NA35,NA35_SS} and the solid curve
is the prediction of the DSP model developed here.  }
\label{fig_SS_prot_rap}
\end{minipage}
\end{figure}

As a first application to nucleus-nucleus collisions, the dynamical string-parton model
is applied to the interaction $S+S$ at $200$ A GeV/c.
The results of the  calculations are to be compared to CERN
data \cite{NA35,NA35_SS} for central S+S reactions. As was the case for $p\,$--$A$ collisions
the calculations are completely predictive; all of the parameters of the model have been
fixed by $e^+e^-$ and $p\,$--$p$ data. In Fig.~\ref{fig_SS_neg_rap} the rapidity distribution 
for negatively charge hadrons is presented. For a pure prediction, this is a quite nice 
result.  The theory predicts the midrapidity region quantitatively correctly. The only 
discrepancy is that the theory overpredicts the production of particles at the larger rapidity 
by a small amount. This could be a result of not having quite enough stopping in the theory. 
In Fig.~\ref{fig_SS_prot_rap} we show the rapidity distribution of net protons for the same 
reaction. The results are again quite good. There is not complete stopping for this reaction 
as two identifiable peaks remain, both in the data and in the theory. The theory does predict 
the absence of a sharp proton fragmentation peak and the filling of the midrapidity region. 
\begin{minipage}[thbp]{6.5in}
\begin{table}[thbp]
\caption
{\footnotesize Slope parameters ($(GeV/c)^{-1}$) for the $p_T$ distributions 
of the final state pions and kaons
in $p$+$S$ and $p$+$Au(Pb)$ collisions at 200 GeV per nucleon. The experimental
values are from Ref.~\protect\cite{NA44}.  
}
\label{table6}
\begin{tabular}{p{1.8cm}p{3.4cm}p{3.4cm}p{3.4cm}p{3.4cm}}
Reaction &  \quad  Pions      & \quad  Kaons          & \quad Pions   &  \quad Kaons  \\
         &  Experiment (NA44) &  Experiment (NA44)    &  DSP Model    &   DSP Model    \\
\hline	   
$p+S$       & \quad $139\pm 3$  &\quad $160 \pm 12$   & \quad $135$   &\quad $141$   \\ 
$p+Au(Pb)$  & \quad $145\pm 3$  &\quad $152 \pm 5$    & \quad $142$   &\quad $145$   \\       
$S+S$       & \quad $154\pm 5$  &\quad $163 \pm 5$    & \quad $147$   &\quad $143$   \\ 
\end{tabular}
\end{table}
\end{minipage}
The data would allow for a small increase in 
stopping which would fill in the midrapidity 
region slightly more, where the theory is within
the error bars but is consistently low. We 
saw for $p\,$--$A$ collisions in Figs.\ref{pS_fig1}-\ref{pAu_fig2}, particularly for $p+Au$, 
that the model was not producing quite as much stopping as is contained in the data. This is 
less noticeable here, but remains a consistent interpretation of the comparison of theory to 
data. 

\vspace*{0.5cm}
 
\begin{figure}[thbp]
%\vspace*{-10mm}
\begin{minipage}[t]{7.5cm}
{\epsfxsize=7cm\epsfysize=4.7cm \epsfbox{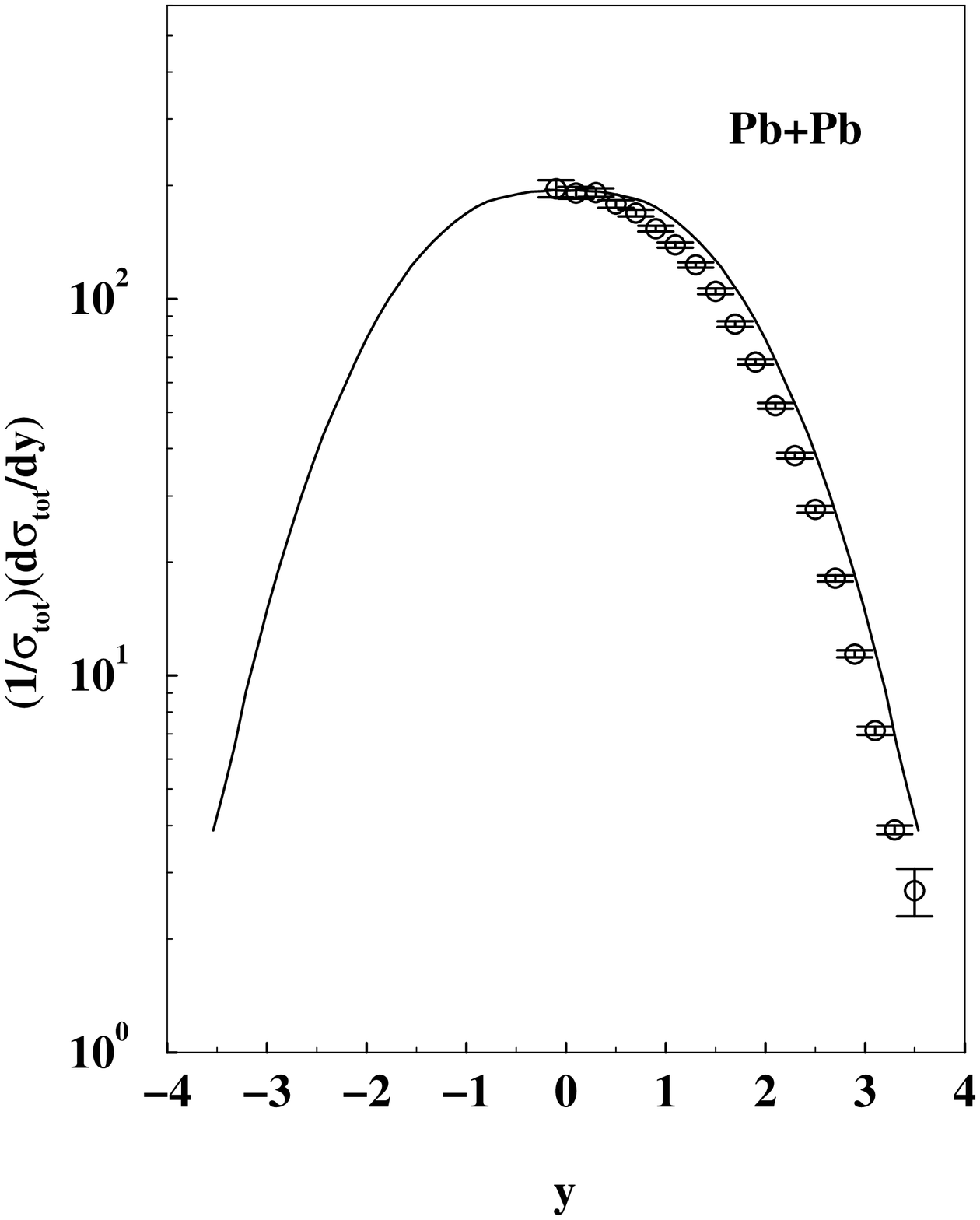}}
\caption[]{\footnotesize Rapidity distribution of negatively charged hadrons
for central $Pb$+$Pb$ collisions at 
156 GeV per nucleon. The data are from \protect\cite{NA49} and the solid curve
is the prediction of the DSP model developed here.  }
\label{fig_Pb_neg_rap}
\end{minipage} \hfill
\vspace*{-7.2cm}
 \hspace*{6.5cm} \hfill
\begin{minipage}[t]{7.5cm}
{\epsfxsize=7cm\epsfysize=4.7cm \epsfbox{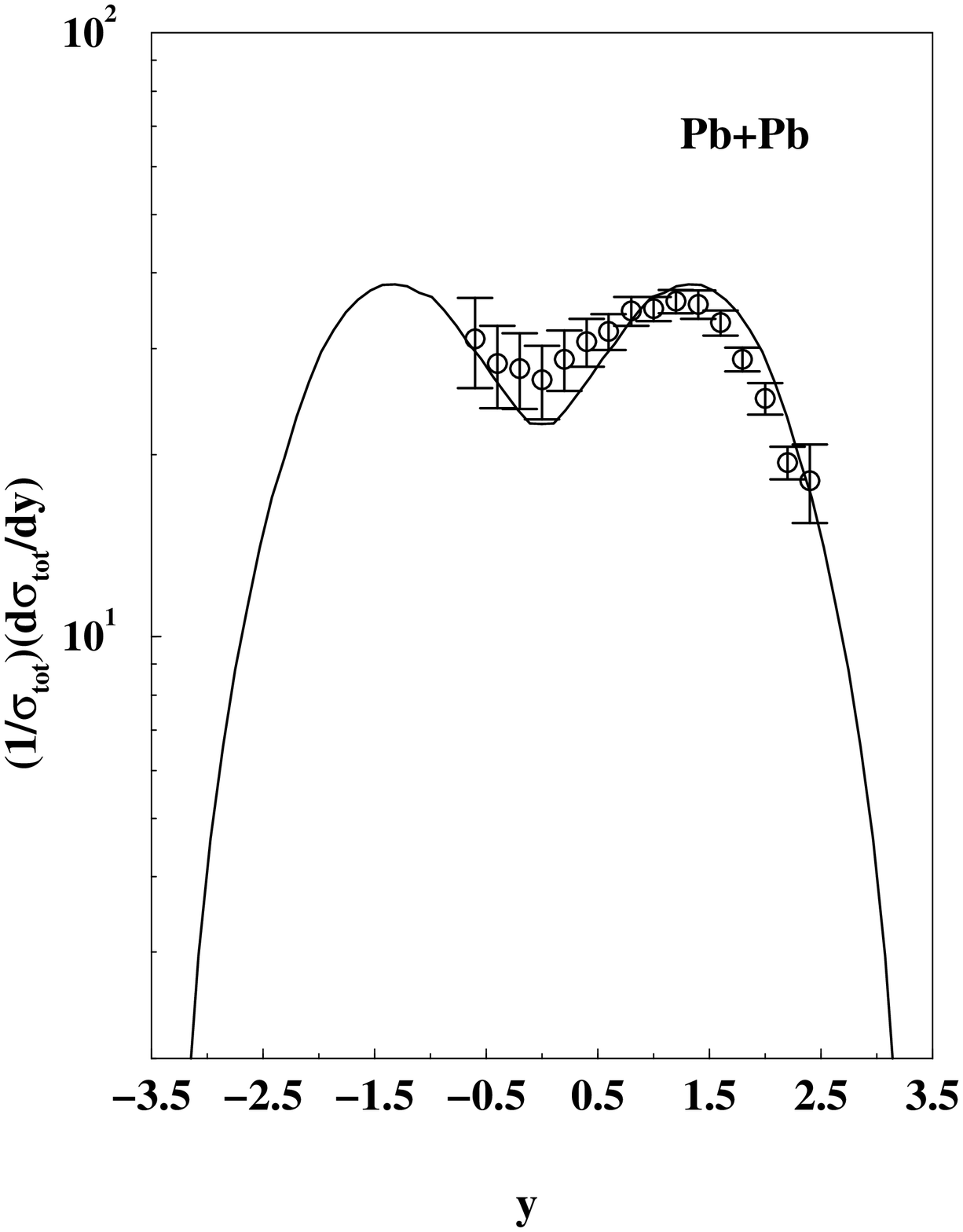}}
%\vspace*{2pt}
\caption[]{\footnotesize  Rapidity distribution of  net protons
for central $Pb$+$Pb$ collisions at 
156 GeV per nucleon. The data are from \protect\cite{NA49} and the solid curve
is the prediction of the DSP model developed here.   }
\label{fig_Pb_prot_rap}
\end{minipage}
\end{figure}

In Fig.~\ref{fig_Pb_neg_rap} we present the distribution of negatively charged particles for 
the collision $Pb+Pb$ at 156 A GeV/c. The data \cite{NA49} are from CERN. Comparing to the 
$S+S$ results
\hspace*{-7mm} 
 in Fig.~\ref{fig_SS_neg_rap}, the results are very comparable. The midrapidity, 
$y=0$, value is predicted very well. For the largest rapidity, the theory is again 
overpredicting the experimental results but not by a large amount. In 
Fig.~\ref{fig_Pb_prot_rap} the rapidity distribution of net protons for the $Pb+Pb$ reaction 
is shown. Again, the results are very similar to what was found for $S+S$ in 
Fig.~\ref{fig_SS_prot_rap}.  There is again incomplete stopping for this reaction as two 
identifiable peaks remain, both in the data and in the theory. The theory does predict the 
absence of a sharp proton fragmentation peak and the filling of the midrapidity region. Again, 
the data would allow for a small increase in stopping which would fill in the midrapidity 
region somewhat more. In summary, the theoretical rapidity distributions for $A$--$A$ 
collisions at CERN energies are quite reasonable, but indicate that the model is slightly 
underpredicting the stopping. 

\begin{figure}[thbp]
{\epsfxsize=7cm\epsfysize=4.7cm \epsfbox{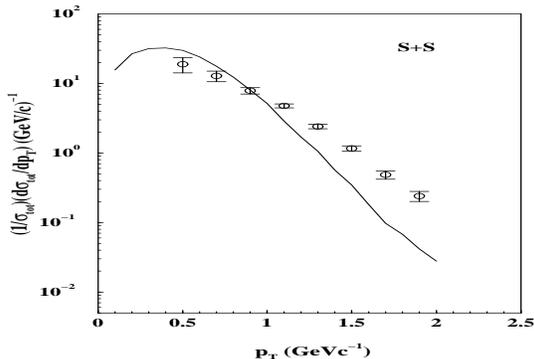}}
%\vspace*{2pt}
\caption[]{\footnotesize  Transverse momentum distributions of net protons 
for central  $S+S$ collisions at 
200 GeV per nucleon. The data are from \protect\cite{NA35} and the solid curve
is the prediction of the DSP model developed here.   }
\label{fig_SS_prot_pT}
\end{figure}

In Fig.~\ref{fig_SS_prot_pT} we present the transverse momentum distribution of net protons 
for the $S+S$ reaction at 200 A GeV/c and the experimental results from \cite{NA35}.
The theory is  underpredicting somewhat the creation of transverse momentum for the protons.
We present in 
\vspace*{0.5cm}
\hspace*{-7mm}
Table \ref{table6}the inverse slope parameters for transverse momentum 
distributions for pions and kaons. 
The reactions are $p+S$, $p+Au(Pb)$, and $S+S$ at 200 A 
GeV. In $A$--$A$ collisions, there is an increase in transverse momentum production in the 
data that is partially found in the theory. The experimental  inverse slope parameter for 
pions produced in $p+S$ collisions were found to be 
$139 \pm 3$ MeV  and $145 \pm 3$ MeV for $p+Au$. The DSP model predicts 135 MeV and 142 MeV 
respectively, results that reasonably consistent. For $S+S$ collisions the experimental 
inverse slope for pion production is $154 \pm 5$ MeV while the DSP model prediction is 147 
MeV, which is just outside the error bars.  The transverse momentum production for kaons is 
consistently underpredicted by a small amount in the DSP model. The experimental inverse 
slopes are $160 \pm 12$ MeV for $p+S$ and $152 \pm 5$ for $p+Au$ . The theoretical results are 
135 MeV and 142 MeV, respectively. Both of these theoretical numbers are somewhat small. For 
$S+S$ collisions the inverse slope parameter for kaon production is $163 \pm 5$ MeV for the 
data and 143 MeV for the DSP model. This data for kaon production
for the $S+S$ reaction indicates a rather large increase in the inverse slope over  that for 
$p$+$A$ collisions, something not found in the model. 

\section{Conclusions} 

In this paper, we further developed the
string parton model of relativistic heavy-ion collisions of Ref.~\cite{DEAN's,Big,LNS,HM}.
This dynamical model is based on Nambu-Got\=o classical 
strings \cite{artru74}. The strings evolve in time according to their covariant equation of 
motion and the endpoints of the string, which contain a finite amount of energy-momentum, are 
identified as quarks, antiquarks, or diquarks.  Hadronization proceeds through a covariant and 
stochastic model of string breaking in which the probability of breaking is proportional to 
the invariant area swept out by the string. The quark-antiquark pair created at the point 
where the string breaks acquire a transverse momentum taken from an exponential distribution. 
The interaction between particles is modeled as an elastic scattering between the massless 
quarks. Previous work \cite{DEAN's,Big,LNS,HM} has found that the model reproduces quite well 
the average properties of $p\,$--$A$ and $A$--$A$ collisions.

In order to address more detailed properties of hadronic interaction data, we extended
the model.  First, the masses of the strings are phenomenologically quantized \cite{DM98} to 
allow for particle identification in the final state. This is done by choosing windows such 
that if the invariant mass of a string falls into an energy region, the string is identified 
with the baryon of that mass and no longer allowed to decay. The incident nucleons are taken 
to be a distribution of string solutions, with the distribution determined \cite{DM99} from 
the experimentally measured nucleon structure function. The gluon content of the nucleon is 
also included by adding gluonic strings where the distribution of string shapes is  determined 
by the gluonic structure function. 
The interaction between quarks and gluons is also generalized to include a hard component, 
taken from perturbative QCD, and a soft component, determined phenomenologically from $e^+e^-$ 
and $p\,$--$p$ collisions. The partons interact via the hard (soft) interaction if their 
center-of-mass energy is above (below) a chosen threshold. In addition, we allow the quarks 
which are created by a string breaking to interact with each other if they are within an 
interaction radius when they come on-shell.

There are a limited number of parameters in the model, and all are determined from $e^+e^-$ 
and $p\,$--$p$ data. These are given in Tables \ref{table1} and \ref{table5}. The model is 
completely predictive at the $p\,$--$A$ and $A$--$A$ level. We present here the first 
calculations of the extended model. For $e^+e^-$ and $p\,$--$p$ collisions we find excellent 
results. This is partially because we adjust our parameters to fit this data. However, 
we find that over the energy range we examined, the 
limited number of energy-independent parameters used does reproduce well a broad range of 
data. We then calculate results for $p\,$--$A$ and $A$--$A$ collisions at CERN energies. On 
the whole, the results are quite satisfactory. The predictions generally lie within the 
experimental error bars with a few points lying just outside the error bars, as would be 
expected. We see perhaps more stopping in the data than is 
produced by the DSP model and an indication of additional transverse momentum production in 
the $S+S$ collision. 

There is considerable work to be done with the model. The model can predict the energy 
dependence of distributions measured in $A$--$A$ collisions. Predictions of these quantities 
up to and including RHIC energies are possible. A comparison with the data that will soon be 
coming forth from RHIC will provide insight into the dynamics of these complex collisions. We 
have calculated only several measured distributions. There are a number of other quantities 
which are both measurable and are calculable by the DSP model. We expect that the model 
will need further refinement as more detailed properties of the reaction are confronted.
Finally, a distinct advantage of the model is that it is dynamic; the reaction is generated as 
a function of time on the computer. This allows us to track in detail the time evolution of 
important aspects of the reaction, such as the central energy density or the flow of energy or 
matter. Understanding the time evolution of the reaction can provide great insight on the 
origin of the asymptotic properties that are to be experimentally measured.

The ultimate goal of relativistic heavy-ion research 
is to discover and investigate the quark gluon plasma. This will to a great 
extent be an exercise in understanding the differences between model predictions and data. It 
is thus very important to constrain the models. In the DSP model, the constraints come from 
$e^+e^-$, $p\,$--$p$, and $p\,$--$A$ data. There exists a limited amount of data for the 
$p\,$--$p$ and $p\,$--$A$ reactions and this data has reasonably large errors. Future data on 
these reactions would provide additional constraints on the DSP model as well as other models.

\acknowledgements  The work of DEM, ASU, and DJE  was supported in part by the US 
Department of Energy under grant DE-FG02-96ER40975. Oak Ridge National Laboratory is 
managed by UT-Battelle, LLC for the US Department of Energy under Contract No. 
DE-AC05-00OR22725.


\begin{thebibliography}{99}
\bibitem{comment}
   The data from Ref.~\protect\cite{NA35} also include a point at the incident target 
   rapidity. This point is not a directly measured point but is inferred from the data to 
   account for nucleons from the target that do not participate in the reaction but 
   continue on with the incident momentum. We have not performed such an analysis and 
   thus do not include this point in our figures.
\end{thebibliography}
\end{document}